\RequirePackage{rotating}
\documentclass[useAMS,usenatbib]{mnras}

\usepackage{ulem}
\usepackage{color}
\usepackage{graphics,graphicx}
\usepackage{times} 
\usepackage{amssymb}
\usepackage{amsmath}
\usepackage{natbib}
\usepackage{lscape}
\usepackage{url}
\usepackage{multirow}
\usepackage{ctable}
\usepackage{threeparttable}
\usepackage[labelfont=bf,labelsep=period]{caption}
\newif\ifAMStwofonts
\AMStwofontstrue

\def\rosat{{\it ROSAT}}

\def\xmm{{\it XMM-Newton}}

\def\suzaku{{\it Suzaku}}

\def\chandra{{\it Chandra}}
\def\swift{{\it Swift}}
\def\swiftng{{\it Neil Gehrels Swift Observatory}}

\def\epicpn{{EPIC-pn}}
\def\epicmos1{{EPIC-MOS1}}
\def\epicmos2{{EPIC-MOS2}}
\def\epicmos{{EPIC-MOS}}

\def\nustar{{\it NuSTAR}}

\def\xrism{{\it XRISM}}
\def\athena{{\it Athena}}
\def\erosita{{\it eROSITA}}

\def\hzps{Hz~s$^{-1}$}

\def\kmps{\hbox{$\rm\thinspace km~s^{-1}$}}
\def\pcmsq{\hbox{$\rm\thinspace cm^{-2}$}}

\def\H0{{km~s$^{-1}$~Mpc$^{-1}$}}

\def\ctps{\hbox{$\rm\thinspace ct~s^{-1}$}}

\def\ergpcmsqps{\hbox{$\rm\thinspace erg~cm^{-2}~s^{-1}$}}

\def\ergps{\hbox{erg~s$^{-1}$}}

\def\msun{\hbox{$M_{\odot}$}}

\def\addascaspec{\textsc{addascaspec}}

\def\flx2xsp{\textsc{flx2xsp}}

\def\sas{\textsc{sas}}
\def\xmmselect{\textsc{xmmselect}}

\def\epchain{\textsc{epchain}}
\def\emchain{\textsc{emchain}}

\def\rmfgen{\textsc{rmfgen}}
\def\arfgen{\textsc{arfgen}}

\def\hendrics{\textsc{hendrics}}

\def\chisq{{$\chi^{2}$}}

\def\xspec{\hbox{\small XSPEC}}

\def\tbabs{\textsc{tbabs}}

\def\diskbb{\textsc{diskbb}}
\def\diskpbb{\textsc{diskpbb}}

\def\eg{{\it e.g.}}
\def\etc{{\it etc.}}
\def\ie{{\it i.e.~\/}}

\def\la{\mathrel{\hbox{\rlap{\hbox{\lower4pt\hbox{$\sim$}}}{\raise2pt\hbox{$<$}}}}}
\def\ga{\mathrel{\hbox{\rlap{\hbox{\lower4pt\hbox{$\sim$}}}{\raise2pt\hbox{$>$}}}}}

\def\d25{D$_{25}$}

\def\.25{0.25 keV\thinspace}

\def\kbol210{\rm $\kappa_{2-10}$}

\def\NS{$N(>$\textit{S}$)$}
\def\nulx{1843}
\def\nulxnew{689}
\def\nulxrecent{1318}
\def\nulxallmiss{50}
\def\nhostgals{951}
\def\nhostgalsmult{333}
\def\fracspiral{$\sim$60}
\def\nhlx{71}
\def\nhlxnew{48}
\def\nhlxrecent{59}

\title[A New Multi-Mission ULX Archive]{A Multi-Mission Catalogue of Ultraluminous X-ray Source Candidates}

\author[D.\,J. Walton et al.]
{\parbox{7.in}{D.\,J. Walton$^{1,2}$ \thanks{E-mail: dwalton@ast.cam.ac.uk},
A. D. A. Mackenzie$^{3}$, 
H. Gully$^{4}$, 
N. R. Patel$^{5}$, 
T. P. Roberts$^{3}$,
H. P. Earnshaw$^{6}$, \\ 
S. Mateos$^{7}$ 
\\[0.25cm]
\footnotesize
$^{1}$ \it{Centre for Astrophysics Research, University of Hertfordshire, College Lane, Hatfield AL10 9AB, UK} \\
$^{2}$ \it{Institute of Astronomy, University of Cambridge, Madingley Road, Cambridge CB3 0HA, UK} \\
$^{3}$ \it{Centre for Extragalactic Astronomy \& Department of Physics, Durham University, Department of Physics, South Road, Durham DH1 3LE, UK} \\
$^{4}$ \it{School of Physics, Astronomy, University of Nottingham, University Park, Nottingham NG7 2RD, UK} \\
$^{5}$ \it{Department of Physics and Astronomy, Pevensey Building, University of Sussex, Brighton BN1 9QH, UK} \\
$^{6}$ \it{Cahill Center for Astronomy and Astrophysics, California Institute of Technology, Pasadena, CA 91125, USA} \\
$^{7}$ \it{Instituto de F\'{i}sica de Cantabria (CSIC-UC), Avenida de los Castros, 39005 Santander, Spain} \\
}}
\date{}

\begin{document}
\pagerange{\pageref{firstpage}--\pageref{lastpage}}
\maketitle
\label{firstpage}

\begin{abstract}
We present a new, multi-mission catalogue of ultraluminous X-ray source (ULX)
candidates, based on recent data releases from each of the \xmm, \swift\ and \chandra\
observatories (the 4XMM-DR10, 2SXPS and CSC2 catalogues, respectively). This has
been compiled by cross-correlating each of these X-ray archives with a large sample of
galaxies primarily drawn from the HyperLEDA archive. Significant efforts have been
made to clean the sample of known non-ULX contaminants (e.g. foreground stars,
background AGN, supernovae), and also to identify ULX candidates that are common to
the different X-ray catalogues utilized, allowing us to produce a combined `master' list
of unique sources. Our sample contains \nulx\ ULX candidates associated with
\nhostgals\ different host galaxies, making it the largest ULX catalogue compiled to date.
Of these, \nulxnew\ sources are catalogued as ULX candidates for the first time. Our
primary motivation is to identify new sources of interest for detailed follow-up studies,
and within our catalogue we have already found one new extreme ULX candidate that
has high S/N data in the archive: NGC\,3044 ULX1. This source has a peak luminosity of
$L_{\rm{X,peak}} \sim 10^{40}$\,\ergps, and the \xmm\ spectrum of the source while at
this peak flux is very similar to other, better-studied extreme ULXs that are now
understood to be local examples of super-Eddington accretion. This likely indicates that
NGC\,3044 ULX1 is another source accreting at super-Eddington rates. We expect that
this catalogue will be a valuable resource for planning future observations of ULXs --
both with our current and future X-ray facilities -- to further improve our understanding
of this enigmatic population.
\end{abstract}


\section{Introduction}

Our understanding of ultraluminous X-ray sources (ULXs) -- off-nuclear X-ray sources
with luminosities in excess of 10$^{39}$\,\ergps\ -- has evolved significantly over the
past few years. Historically, the debate regarding the nature of these sources has
focused on whether they represent a population of sub-Eddington `intermediate mass'
black holes (IMBHs with $M_{\rm{BH}} \sim 10^{2-5}$\,\msun; \eg\ \citealt{Colbert99,
Miller04, Strohmayer09a}) or a population of super-Eddington but otherwise normal
stellar remnants (\eg\ \citealt{King01, Poutanen07, Middleton15}); see \cite{Kaaret17rev}
for a recent review. Although evidence for large black holes is now being seen by LIGO
(most notably the recent detection of a BH--BH merger resulting in a $\sim$150\,\msun\ 
remnant; \citealt{LIGO20_IMBH}), the general consensus is now that the majority of
ULXs represent a population of super-Eddington accretors, thanks in particular to the
broadband spectral and timing studies possible in the \nustar\ era (\citealt{NuSTAR})
and the high-resolution spectra provided by \xmm\ (\citealt{XMM}).

The broadband spectra obtained early in the \nustar\ mission demonstrated that 
ULX spectra are clearly distinct from standard modes of sub-Eddington accretion (\eg\
\citealt{Bachetti13, Walton13culx, Walton14hoIX, Walton15hoII, Rana15,
Mukherjee15}), confirming prior indications from \xmm\ (\eg\ \citealt{Stobbart06,
Gladstone09}), and instead revealed high-energy spectra consistent with broad
expectations for super-Eddington accretion (\ie spectra that appear to show a strong
contribution from hot, luminous accretion discs, \eg\ \citealt{Shakura73, Abram88,
Poutanen07}). The super-Eddington nature of at least some ULXs was then
spectacularly confirmed with the discovery that the ULX M82 X-2 ($L_{\rm{X,peak}}
\sim 2 \times 10^{40}$\,\ergps) is actually powered by a highly super-Eddington
neutron star, following the detection of coherent X-ray pulsations (\citealt{Bachetti14nat}).
Five more ULX pulsars have since been discovered (\citealt{Fuerst16p13, Israel17p13,
Israel17, Carpano18, Sathyaprakash19, Rodriguez20}), revealing an accretion regime that
extends up to $L/L_{\rm{E}} \sim 500$. In addition to the broadband spectra and the
discovery of ULX pulsars, we now have evidence in ULX data for the powerful outflows
ubiquitously predicted by models of super-Eddington accretion via the detection of
blueshifted atomic features. These have been seen primarily in the low-energy \xmm\
RGS data, but also in the iron K band in a couple of cases (\citealt{Pinto16nat, Pinto17,
Pinto20, Walton16ufo, Kosec18ulx, Kosec18}). These outflows exhibit extreme velocities
($\sim$0.1--0.3$c$), implying they carry a significant additional energetic output from
these already extreme X-ray binary systems.

Nevertheless, important questions still remain regarding the ULX population. Although it
is now speculated that ULX pulsars could actually make up a significant fraction of these
sources (\eg\ \citealt{Pintore17, Koliopanos17, Walton18ulxBB}), their exact contribution
is still highly uncertain. Is there also a significant population of black hole ULXs, and if
so could these be the progenitors of the BH--BH mergers now regularly being seen by
LIGO (\citealt{Inoue16, Mondal20})? Given the history of the field, it is easy to forget that
we still do not have a single ULX with a well-constrained mass function that
unambiguously requires the accretor to be a black hole. Can black hole ULXs (assuming
they exist) reach the same extreme Eddington ratios as ULX pulsars, or is this somehow
related to the magnetic nature of these objects (as suggested by \citealt{DallOsso15,
Mushtukov15})? What fraction of the total energetic output is radiative, and what fraction
is kinetic (\ie carried by winds/outflows) at these extreme accretion rates? Understanding
this last issue may in turn be critical for understanding early-universe SMBH growth (and
associated feedback), given that $\sim$10$^{9}$\,\msun\ black holes are now being
observed when the universe was only $\sim$0.7\,Gyr old (\eg\ \citealt{Banados18nat}).

Furthermore, although the overall population is now expected to be dominated by
super-Eddington accretors, there are still rare individual sources among the ULX
population that remain good IMBH candidates. Most notable among these is the case of
ESO\,243--49 HLX1, which reaches an astonishing luminosity of $L_{\rm{X,peak}} \sim
10^{42}$\,\ergps\ (\citealt{Farrell09}). In contrast to the vast majority of the ULX
population, this source does behave as expected for a scaled-up sub-Eddington X-ray
binary (\citealt{Servillat11, Webb12}). Furthermore, M82 X-1 has long been thought of
as an IMBH candidate because of its X-ray properties (\eg\ \citealt{Feng10,
Pasham14nat}, although see \citealt{Brightman20m82} for caveats), and NGC\,2276--3c
has also been suggested as an IMBH candidate owing to its position on the radio--X-ray
fundamental plane (\citealt{Mezcua15}). Identifying further IMBH candidates remains of
significant interest, given the scarcity of compelling cases among the ULX population.

Key to advancing all of these areas are efforts to grow the broader ULX population and
provide larger samples with which to undertake statistical studies of ULXs and identify
notable individual sources for follow-up study. Most previous efforts have focused on
searching for ULXs in individual mission archives, using in particular \rosat\
(\citealt{Roberts00, Colbert02, LiuBregman05}), \chandra\ (\citealt{Swartz04, Liu11,
Gong16, Kovlakas20}) and \xmm\ (\citealt{WaltonULXCat, EarnshawULXcat}). Focusing
on data from a single mission has the advantage that everything (source selection,
energy bands) can be treated in a uniform manner, which is important for performing
population-based studies where selection biases need to be carefully controlled.
However, this comes at the expense of limiting the sky area/temporal coverage utilized
relative to that available in the full, multi-mission X-ray archive, both of which are key
factors in terms of identifying individual sources that may be of particular interest. 

Here, we present the results of a search for new ULX candidates, combining data from all
of the public archives from the major soft X-ray imaging observatories currently
operational: \xmm\ (\citealt{XMM}), \chandra\ (\citealt{Chandra}) and the \swiftng\
(hereafter \swift; \citealt{SWIFT}). In particular, we make use of the 4XMM-DR10, CSC2
and 2SXPS source catalogues (\citealt{4XMM, CSC2temp, 2SXPS}). Although combining
the data from these facilities does formally introduce some non-uniformity to the
selection, our primary aim is to compile the largest raw sample of ULX candidates to date,
facilitating searches for sources that are bright enough for detailed follow-up with current
and future X-ray facilities, as well as searches for sources with multi-wavelength
counterparts. This is of particular interest with both \xrism\ (\citealt{XRISM_tmp}) and
\athena\ (\citealt{Athena}) on the horizon, as well as the new facilities due to come online
at longer wavelengths (\eg\ thirty-metre class optical telescopes, \textit{JWST}, the SKA,
\etc).

The paper is structured as follows: in Section \ref{sec_gals} we outline the galaxy sample
within which we search for ULX candidates, and in Section \ref{sec_search} we discuss
our procedure for identifying ULX candidates from the individual archives. Section
\ref{sec_sample} presents our final, merged sample of ULX candidates, and we highlight
the case of NGC\,3044 ULX1 -- a new extreme ULX discovered here -- in Section
\ref{sec_ulxs}. Finally, we summarise our findings in Section \ref{sec_conc}.

\section{Galaxy Sample}
\label{sec_gals}

In addition to the various X-ray archives considered here, the other major input required
for this work is a catalogue of galaxies within which to search for ULXs. Here, we primarily
use the HyperLEDA database (\citealt{HYPERLEDA}), initially selecting everything labelled
as a galaxy. We focus on HyperLEDA because this is one of the largest homogenized
compilations of  known galaxies available in the literature. However, we further supplement
these galaxies with the latest version of the Catalogue of Nearby Galaxies (CNG;
\citealt{Karachentsev18}).

For our work here, we need to be able to define the sky area subtended by the galaxy (in
order to positionally match X-ray sources) as well as the distance to the galaxy (in order
to compute source luminosities). For the galaxy areas, we assume the extent of each
galaxy is determined by the elliptical region defined by its D25 isophote (\ie the best
elliptical fit to the area over which $B$-band surface brightness of the galaxy exceeds
25 mag arcsec$^{-2}$), which is given in HyperLEDA (where this information is available).
However, CNG uses the Holmberg radius to define the semi-major axis of the galaxy
ellipse instead, which corresponds to a surface brightness of 26.5 mag arcsec$^{-2}$.
For the subset of galaxies included in both HyperLEDA and CNG, we therefore calculate
an empirical correction between the D25 semi-major axes ($R_{\rm{D25}}$) and the
Holmberg radii ($R_{\rm{Holm}}$), and then apply this to any remaining galaxies that are
only included in CNG in an attempt to normalise these to the D25 definition. On average,
we find $R_{\rm{Holm}} = (1.26 \pm 0.02) R_{\rm{D25}}$ (where the uncertainty quoted
here is the 1$\sigma$ standard error on the mean). Although the full set of D25
information (semi-major axis, semi-minor axis, position angle) is obviously required to
search the full sky area subtended by the galaxy, in cases where the position angle is
missing it is still possible to search for ULX candidates within a circular region defined
by the semi-minor axis, as this will always be within the galaxy area regardless of the
orientation of its full elliptical region. We therefore also retain these galaxies in our input
sample. However, any galaxies in the HyperLEDA/CNG catalogues that do not have
sufficient information that we can compute at least their D25 semi-minor axes are
discarded.

For the majority of the galaxies considered, we compute distances based on their
measured redshifts ($z$) assuming they adhere to the Hubble flow. However, where
redshift-independent distance estimates are available, we prioritise these
measurements. HyperLEDA and CNG both include these based on a variety of different
methods (via \eg\ Cepheid variables, the tip of the red giant branch, the Tully-Fischer
relationship), and we further supplement these with distance measurements from the
latest version of the Cosmicflows galaxy catalogue (\citealt{Tully16}). Such
measurements are particularly critical for very nearby galaxies (recession velocities
$cz < 1000$\,\kmps), where peculiar motions can dominate over the Hubble flow. For
these galaxies, we therefore also collected further redshift-independent distance 
from the NASA Extragalactic Database where such measurements were not available in
any of the HyperLEDA/CNG/Cosmicflows catalogues. Where there are multiple distance
estimates available among these catalogues, we prioritise them in the following order:
Cosmicflows $>$ CNG $>$ Hyperleda $>$ NED $>$ Hubble flow, but we stress that in
the majority of cases there is generally good agreement between the different
catalogues regarding the redshift-independent distance estimates. However, since a
reasonably reliable distance estimate is in turn critical for a reliable luminosity
calculation, we therefore discard galaxies with recession velocities $cz < 1000$\,\kmps\
where there is no redshift-independent distance estimate available in any of the above 
(similar to both \citealt{WaltonULXCat} and \citealt{EarnshawULXcat}).

\begin{figure}
\begin{center}
\hspace*{-0.35cm}
\rotatebox{0}{
{\includegraphics[width=235pt]{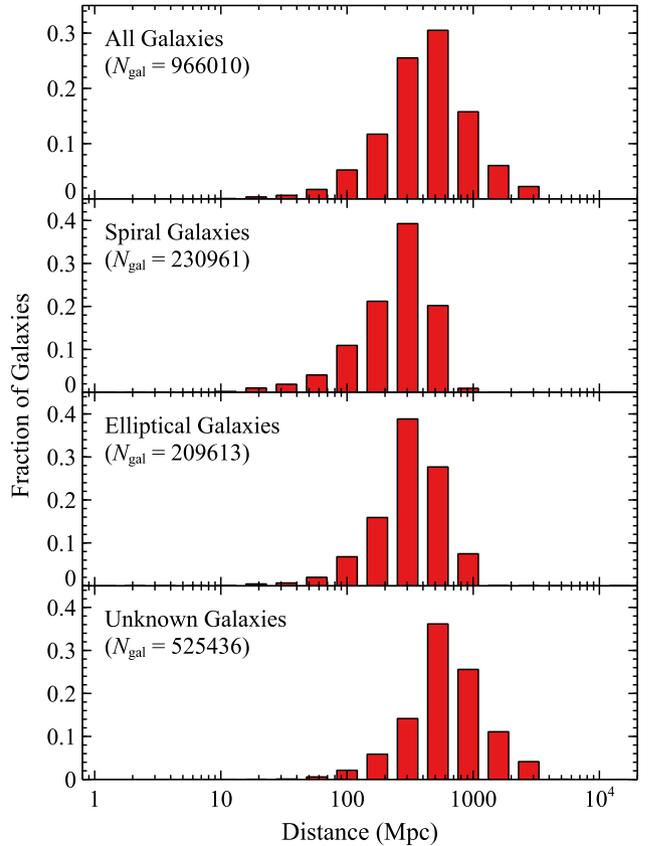}}
}
\end{center}
\vspace*{-0.3cm}
\caption{
Distance distributions for the full galaxy sample utilized here (top panel), galaxies
identified as spiral ($T \geq 1$) and elliptical ($T < 1$) galaxies (upper-middle and
lower-middle panels, respectively), and for galaxies where morphology information is
not available (bottom panel).}
\label{fig_galdist}
\end{figure}

The final galaxy sample utilized here consists of 966,010 entries, after accounting
for the requirements outlined above, the vast majority of which come from HyperLEDA
(only 215 of these galaxies are found in CNG but not HyperLEDA). Just under half of
these galaxies have morphology estimates available in the form of the Hubble type, $T$.
Following \cite{WaltonULXCat}, for these galaxies we make the distinction between spiral
galaxies ($T \geq 1$, including irregular galaxies) and elliptical galaxies ($T < 1$,
including lenticular galaxies). We show the distance distributions for the full galaxy
sample utilized here, as well as some of these subsets, in Figure \ref{fig_galdist}. The
majority of the galaxies considered are within a Gpc, although the galaxies for which
morphology information is not available do have larger distances on average than those
where the morphology has been identified.

\section{Selection of ULX Candidates}
\label{sec_search}

\subsection{Basic Approach}

We take the same basic approach to selecting ULXs for each of the three X-ray source
catalogues utilized here (4XMM-DR10, 2SXPS and CSC2). Our initial analysis of these
individual archives can be broadly summarised into 5 main steps, as described below.
Many of the specific details differ for the different catalogues utilized, owing to the
differences between the different X-ray observatories they are derived from and the
different formats in which the data are provided; these details are discussed in Section
\ref{sec_Xcats}.

\textit{\textbf{Step 1 -- positional match:}}
First, we perform a positional cross-match between our input galaxy list and each X-ray
catalogue; as noted above, for galaxies where the full set of spatial information is available
(both the major and minor axes, and the position angle) we perform a standard elliptical
match around the position of the galaxy (\ie\ utilizing the full sky area it subtends), while
for galaxies where the position angle is missing we perform a circular match within the
radius set by the semi-minor axis (and thus potentially only utilize a fraction of its sky
area). Where relevant, only X-ray sources listed as being point-like are retained (both
CSC2 and 4XMM-DR10 also contain extended sources, while in principle 2SXPS only
includes point sources).

There are inevitably a small subset of X-ray sources that are consistent with being
associated with more than one galaxy in our initial matched source lists. In these cases,
we make sure to retain only one of the repeat entries in order to avoid individual X-ray
sources/detections appearing more than once. To do so, we initially associate the
X-ray source with the galaxy for which it is closest to the centre. This is because we
expect the majority of these cases to be interacting galaxies, which by definition will
be at essentially the same distance, and on average the density of ULXs is known to
increase as you approach the galaxy centre (\citealt{Swartz11}). However, we re-visit
this assumption at the end of our analysis of the individual archives (i.e. after step 5),
and assess whether the different galaxy distances really are similar for any remaining
sources that are potentially associated with more than one galaxy. Here, we treat
galaxies with distances that differ by $<$15\% as having similar distances. Where this
is not the case, such that the potential host galaxies appear to be un-associated
galaxies that happen to overlap when viewed in projection, we switch the association
of the X-ray source to the less distant galaxy. This is both a conservative approach,
resulting in lower X-ray luminosities, and also probably a more realistic assumption in
these situations, as the enhanced absorption by gas and dust in the foreground galaxy
would mean sources in the background galaxy are less likely to be seen as ULXs
based on their observed luminosities. For any sources where the association is
changed at this stage, the luminosities are re-evaluated (see step 2), and any sources
that no longer meet the ULX criterion are excluded.

\textit{\textbf{Step 2 -- luminosity cut:}}
With our positionally matched source lists, we then compute X-ray luminosities using
our preferred galaxy distance. Here, we use the full band fluxes available in each of the
individual X-ray catalogues (see Section \ref{sec_Xcats}). These bandpasses are not
precisely identical, but are sufficiently similar that we consider this a reasonable
compromise for the sake of simplicity, particularly in light of the simple spectral forms
assumed when estimating these fluxes (see Section \ref{sec_Xcats}). Attempting to
correct all of the fluxes to have a common treatment is non-trivial, particularly given
the time-dependent nature of the \chandra\ instrumental responses (owing to the
long-term build-up of the ACIS contamination layer; \citealt{Plucinsky18}). With the
luminosities in hand, we retain only sources with a full band luminosity that exceeds
$10^{39}$\,\ergps, the standard definition of a ULX. In particular, we select sources
that have exhibited luminosities in excess of $10^{39}$\,\ergps\ during any individual
observation of the source, allowing us to select both persistent and transient/highly
variable ULXs. This is a key consideration here, since the latter are now being detected
in increasing numbers as our X-ray archives continue to grow (\eg\ \citealt{Soria12,
Middleton12, Middleton13nat, Pintore18, Pintore20, Earnshaw19, Earnshaw20,
vanHaaften19, Brightman20m51, Walton21}), and may be of particular interest in the
context of identifying good ULX pulsar candidates (\eg\ \citealt{Tsygankov16,
Earnshaw18, Song20}).

\textit{\textbf{Step 3 -- quality flag cut:}}
Each of the X-ray catalogues utilized here contain a variety of information that relate to
the robustness of the X-ray detection included and the source properties derived. In
each case, we make use of this information to ensure that we only consider sources for
which the available X-ray information is reliable, further discarding sources where there
are concerns that this may not be the case. The approach taken here is necessarily
specific to each of the individual catalogues considered, and is detailed below for each
in turn (Section \ref{sec_Xcats}).

\textit{\textbf{Step 4 -- nuclear exclusion:}}
By definition, ULXs are off-nuclear sources, and so we attempt to exclude sources that
may be associated with the nuclei of their host galaxies. However, this is made
challenging by the fact that low-luminosity AGN can exhibit similar luminosities to ULXs
(\citealt{Ghosh08, Zhang09}). We therefore exclude potential AGN through their position
relative to the centre of the galaxy instead, following the approach taken in
\cite{EarnshawULXcat}. In brief, for each X-ray source we compute the minimum
separation from the central galaxy position, $R_{\rm{min}}$, based on its 3$\sigma$
positional uncertainty (\ie\ we define $R_{\rm{min}}$ = nuclear separation $-$
3$\sigma$ position error). We then select sources with $L_{\rm{X}} > 10^{42}$\,\ergps,
as these sources are almost certainly AGN (only one ULX, ESO\,243--49 HLX1, has
exhibited such luminosities to date; \citealt{Farrell09}), and calculate the cumulative
distribution of their $R_{\rm{min}}$ values. Unsurprisingly, these typically exhibit very
small minimum nuclear separations, and we determine the value of $R_{\rm{min}}$ that
contains $>$99\% of these sources, which we take to be our exclusion criterion for
nuclear sources, $R_{\rm{min,excl}}$ (see Section \ref{sec_Xcats}). All sources with
$R_{\rm{min}} < R_{\rm{min,excl}}$ are subsequently excluded from our analysis. We
repeat the assessment of $R_{\rm{min}}$ with both our input galaxy catalogue, where
we have a requirement for a minimum amount of information regarding the extent of
the galaxy (Section \ref{sec_gals}), and also with the full HyperLEDA/CNG/Cosmicflows
galaxy catalogues (as for this stage, only the position of the galaxy is required), such
that sources with $R_{\rm{min}} < R_{\rm{min,excl}}$ for \textit{any} galaxy included in
these databases are excluded from our source lists. This empirical approach allows us
to conservatively account for the uncertainty associated with the fact that for some
galaxies it can be difficult to precisely identify its central/nuclear position (\eg\ irregular
or merging galaxies, and/or offset nuclei).

\textit{\textbf{Step 5 -- removal of other known contaminants:}} 
In addition to the nuclei of the apparent host galaxies for our sources, we also attempt to
remove other known contaminants. At this stage, we particularly focus on background
AGN and foreground stars that coincidentally appear to be associated with a host galaxy
in projection. We therefore positionally match our remaining source lists against lists of
known stars and quasars. For the former, we use the Tycho2 catalogue (\citealt{tycho2})
while for the latter we use the GAIAunWISE quasar catalogue (\citealt{GAIAunWISE}) and
the quasar catalogue of \cite{VeronCetty10}. The search radii we use vary depending on
the X-ray archive, as detailed in Section \ref{sec_Xcats}. Any source that matches with
either a known star or a known quasar is excluded from our analysis.

\subsection{Specific Catalogue Details}
\label{sec_Xcats}

\subsubsection{4XMM-DR10}

The 4XMM catalogue (\citealt{4XMM}) is formatted such that every row entry represents
a unique detection of an X-ray source by the EPIC detectors (pn, MOS1, MOS2;
\citealt{XMM_PN, XMM_MOS}), meaning that the observation-by-observation information
needed to determine the peak flux for sources that have been observed on multiple
occasions is already incorporated. For the initial position match (step 1), we specifically
use the \textsc{ra\_sc} and \textsc{dec\_sc} columns in the 4XMM catalogue for the X-ray
position, which give the catalogue-averaged position for sources that have been detected
on multiple occasions. 4XMM includes both point sources and extended sources, and we
exclude observations in which the detection is marked as extended\footnote{Note that we
primarily use the observation-by-observation measures of source extent here, as
opposed to the mission averaged measure (\textsc{sc\_extent}), as variable/transient
point sources embedded in diffuse emission can still mistakenly be flagged as extended
in the latter. As such, there are some sources included that have non-zero values for
\textsc{sc\_extent}, but we stress that we have inspected the \xmm\  images from the
observations listed as being point-like for these sources that are included in our final
sample, and visually confirmed the presence of a point-like source.}. When computing
detection luminosities (step 2), we use the full band flux provided in the catalogue (i.e.
4XMM band 8, spanning 0.2--12.0\,keV); these fluxes are computed by summing the
fluxes of the 4XMM sub bands, which are themselves computed assuming a standard
spectral shape (an absorbed powerlaw continuum with $\Gamma = 1.7$ and $N_{\rm{H}}
= 3 \times 10^{20}$\,\pcmsq).

For the quality flags (step 3), we largely follow the approach taken in
\cite{EarnshawULXcat}. In brief, detections with a summary flag $\geq$ 2 are excluded
to reduce spurious detections in general, and sources with the out-of-time event flag
(Flag 10) set and a total count rate $<$ 0.05\,\ctps\ are also excluded as these are likely
to be artefacts of out-of-time events that are associated with a nearby bright source.
However, in addition to these cuts, we also filter out entries where the \textsc{maskfrac}
flag (Flag 1) is set to be true for each of the EPIC detectors that registered the detection.
This helps to further limit spurious detections seen at chip edges, and also spurious
`new' detections at the edge of the field-of-view (FoV) that are really associated with
known bright sources just outside the FoV. When filtering out sources consistent with
being the nuclei of the host galaxies (step 4) and identifying likely matches with known
foreground stars/background quasars, we again make sure to use the \textsc{ra\_sc}
and \textsc{dec\_sc} columns for the X-ray positions. For the former, we find
$R_{\rm{min, excl}} = 9''$ following the empirical approach described above. This is a
pretty conservative cut, compared to previous works involving \xmm\ data
(\citealt{WaltonULXCat, EarnshawULXcat}). In the latter case, since we are simply
matching point-source positions, we use a matching radius for the various star/quasar
catalogues of 5$''$, roughly corresponding to the typical $3\sigma$ positional accuracy
for point sources in 4XMM-DR10 (\citealt{4XMM}).

\subsubsection{2SXPS}

By definition the 2SXPS catalogue (\citealt{2SXPS}) only includes point sources detected
by the XRT (\citealt{SWIFT_XRT}), and the main table of the catalogue is formatted such
that every row entry represents a unique X-ray source, with the
observation-by-observation detection information contained in a separate table. However,
the primary source table includes information on the peak flux seen by the XRT for each
source included, and so we mainly use this table for our analysis. As such, for the
initial position match (step 1), we are naturally using the best-fit position determined from
all of the available observations of a given source. When computing the relevant source
luminosity (step 2), we primarily select sources based on the peak flux given for each
source in the full XRT band (spanning 0.3--10.0\,keV) assuming again an absorbed
powerlaw continuum (fluxes for various potential spectral models are provided, but of
these the absorbed powerlaw is the most appropriate choice for ULXs below 10\,keV).
Here, the powerlaw parameters adopted when computing the catalogued fluxes are
either fit directly, derived from the 2SXPS hardness ratios, or a photon index of
$\Gamma = 1.7$ and the Galactic column in the direction of the source are assumed
(see \citealt{2SXPS} for details).

In contrast to both \xmm\ and \chandra, typical \swift\ observations are very short
exposures ($\sim$1--2\,ks). Furthermore, these observations themselves are often split
up into several shorter `snapshots', and the peak flux included in the catalogue can in
principle be drawn from the count rate seen during one of these snapshots instead of
the full observation. As such, the peak flux often has large uncertainties, being based on
only a handful of counts. For sources where the peak luminosity has a fractional error of
$>$40\% (averaging the positive and negative errors quoted), corresponding to a
detection with $\sim$10 counts based on the approximation for the Poisson distribution
presented in \cite{Gehrels86}, we therefore also require that the source meet at least
one of three additional criterion for inclusion. Either: 1) the average luminosity is also
consistent with the ULX regime, assuming that the average and peak luminosities are not
identical, or 2) there are two or more independent detections of the source in the ULX
regime, based on the observation-by-observation data, or 3) the source has also
previously been detected in the ULX regime by some other facility (\ie\ the detection is
spatially consistent with an entry in one of the archival ULX catalogues we compare our
new dataset against; see Section \ref{sec_prev_ulxcat} for further discussion). For the
quality flag cut (step 3), we exclude sources with the summary flag set to $\geq$1 (\ie
the ``clean" criterion defined by the \swift\ team). When filtering out sources consistent
with being the host galaxy nuclei (step 4), we also find $R_{\rm{min, excl}} = 9''$, similar
to our analysis of 4XMM. Finally, when identifying likely matches with known foreground
stars/background quasars (step 5), we use a matching radius of 10$''$ for the \swift\
data, again corresponding to the typical $3\sigma$ positional accuracy for sources in
2SXPS (\citealt{2SXPS}).

\subsubsection{CSC2}
\label{sec_csc2}

Similar to 2SXPS, the main table of the CSC2 catalogue (\citealt{CSC2temp}) is formatted
such that every row entry represents a unique X-ray source, with the
observation-by-observation detection information presented in a separate table, and
similar to 4XMM both point-like and extended sources are included. We therefore use the
primary source table when performing the initial position match with our galaxy catalogue
(step 1), such that we are again using the best-fit position determined from all of the
available observations of a given source, but we then compile the
observation-by-observation information for each matched source from the secondary
table so that we can be sure to account for the peak flux seen by \chandra\ for each
source in our analysis. Sources listed as being extended are discarded. When computing
detection luminosities (step 2), we use the broadband CSC2 fluxes, \ie `broad' fluxes for
the ACIS detectors (\citealt{CHANDRA_ACIS}), spanning 0.5--8.0\,keV, or `wide' fluxes
for the HRC (\citealt{CHANDRA_HRC}), spanning 0.2--10.0\,keV. Where possible we
again use fluxes derived assuming a powerlaw spectral form (as with 2SXPS, fluxes for
a variety of spectral models are provided, see the CSC2 documentation for details). Here
the catalogued powerlaw fluxes we use are computed assuming $\Gamma = 2$ and the
appropriate Galactic column. However, if this powerlaw flux is not available then we use
the raw aperture flux instead. For the quality flag cut, we only consider sources which
are flagged as `true' detections in the primary source table (\ie sources flagged as
`marginal' detections are excluded), and we also exclude source detections at the
observation level for which the `streak' flag is set. 

In addition to the standard filtering steps outlined above, one further issue that is of
relevance for the observation-by-observation \chandra\ data is the fact that the
\chandra\ PSF degrades rapidly with off-axis angle (in a relative sense, much more
severely than is the case for either \xmm\ or \swift). As such, the typical extraction radii
used in CSC2 also increase with increasing off-axis angle; for example, sources with
off-axis angles of $\sim$8$'$ often have extraction radii of $\sim$9--10$''$,
significantly larger than the on-axis PSF. Unfortunately, for point sources that are either
in crowded regions or are embedded in extended regions of diffuse emission, this can
result in spurious fluxes for any significantly off-axis observations, as these off-axis
detections can occasionally be blends of multiple point sources, and/or include
significant diffuse flux not actually associated with the point source in question. This
is particularly an issue for observations of nearby giant elliptical galaxies; \chandra\
has undertaken significant programs tiling a number of these galaxies (\eg\ M87)
resulting in a combination of on- and off-axis observations of the same crowded fields.
As such, there are a number of sources in these galaxies which have very modest
luminosities when viewed on-axis ($L_{\rm{X}} < 10^{38}$\,\ergps) but which all appear
to share the same ULX-level off-axis detection. Although some ULXs can be highly
variable, as noted above, in many of these cases the ULX-level detections are
unfortunately spurious. We therefore manually inspect cases where the only ULX-level
detection is taken significantly off-axis, and there is also an on-axis observation that
shows the source to have a significantly lower luminosity. Where these are clearly
cases relating to source confusion, we exclude these sources from our analysis. We
also exclude cases in which the size of the aperture increases to the point where it
covers the nominal position of the galaxy centre. In cases where the higher flux could
plausibly be due to variability (\ie\ the on-axis observations show no evidence for large
numbers of sources or diffuse emission whose integrated flux could explain that seen
in the off-axis observation) we retain these detections, but stress that they should be
considered high-priority for further (triggered) follow-up to confirm their nature. We
also retain cases in which the aperture marginally overlaps with the edge of the nuclear
exclusion zone (but not the nominal nuclear position).

When filtering out sources consistent with being the host galaxy nuclei (step 4), we find
$R_{\rm{min, excl}} = 6.1''$ for CSC2, smaller than the equivalent value for both 4XMM
and 2SXPS. Although in a qualitative sense this is not surprising, given the superior
imaging capabilities, it is still worth noting that this value is still significantly larger than
the \chandra\ point spread function. This likely reflects the fact that for more complex
galaxy morphologies it can be difficult to accurately identify the position of the true
galaxy centre. Finally, when identifying likely matches with known foreground
stars/background quasars (step 5), we use a matching radius of 3$''$.

\subsection{Merging and Further Filtering}

Once all of the individual catalogues of ULX candidates from each of the 4XMM-DR10,
CSC2 and 2SXPS archives have been produced, we merge them all into a final
`master' ULX catalogue. To do so, we sequentially match our individual \xmm, \swift\
and \chandra\ catalogues of ULX candidates by position. We begin by matching the
\xmm\ and \swift\ catalogues. As \swift\ is typically the limiting factor regarding position
uncertainties, we match the two within a radius of 10$''$, corresponding to the typical
3$\sigma$ 2SXPS position uncertainty. There are three main outcomes from this initial
match: sources with only \xmm\ data, sources with only \swift\ data, and matched
sources with both. Each of these lists are then matched with the \chandra\ catalogue.
For the sources with only \swift\ data, we again use a matching radius of 10$''$ here,
and for the sources with only \xmm\ data we use a matching radius of 5$''$ (again, the
typical 3$\sigma$ 4XMM-DR10 position error, as \xmm\ is the limiting factor regarding
position uncertainties here). For sources with both \xmm\ and \swift\ data, we assume
the \xmm\ position to be more accurate, and so use this to match to the \chandra\
catalogue, again using a matching radius of 5$''$.

At each of these matching stages, there is the possibility that there are multiple
matches for the same source. This is particularly the case when matching either the
\xmm\ or the \swift\ data with \chandra, given the potential for source confusion and
the superior imaging capabilities of the latter; a famous example is the case of
NGC\,2276, in which a source perceived to be an extremely luminous ULX by \xmm\
is actually resolved into three distinct point sources by \chandra\ (\citealt{Sutton12}).
In that case, all of the resolved sources are themselves ULXs, but it is also possible
that a source that appears as a ULX to \xmm\ or \swift\ will actually be resolved into
multiple sub-ULX sources (this is conceptually similar to the issue regarding the
degradation of the off-axis \chandra\ PSF discussed in Section \ref{sec_csc2}). In
addition to matching them against our \chandra\ catalogue of ULX candidates, we
therefore also match our \xmm\ and \swift\ ULX candidates against the set of
\chandra\ sources that did not make our luminosity cut, and again manually inspect
all cases of multiple matches in order to identify sources that only appear to be ULXs
because of a detection that is actually likely the blend of several point sources,
artificially inflating its apparent flux. As before, these sources are removed from our
analysis. We note, however, that we still retain cases where \eg\ \chandra\ resolves an
\xmm\ detection into two discrete sources, but that the \xmm\ data imply that at least
one of these must have varied into the ULX regime (for example, a scenario in which
\chandra\ sees two sources at $L_{\rm{X}} \sim 10^{38}$\,\ergps, but the \xmm/\swift\
detection that is consistent with both of these sources implies $L_{\rm{X}} \sim 2
\times 10^{39}$\,\ergps, meaning that at least one of these sources must have been in
the ULX regime during the \xmm/\swift\ observation). In these cases, we add a flag to
the final version of the master catalogue noting that this issue exists (a separate flag is
added for each of the matched catalogue pairs, see Table \ref{tab_matchflag} for the
definitions of the different values these flags can take).

On occasion, where there are multiple matches it is possible to determine with
reasonable confidence which of the resolved sources the unresolved detection is
actually associated with (for example, cases where \chandra\ sees two sources, one
with $L_{\rm{X}} \sim 10^{38}$\,\ergps and one with $L_{\rm{X}} \sim 10^{40}$\,\ergps\
and \xmm/\swift\ sees one source that also has $L_{\rm{X}} \sim 10^{40}$\,\ergps, or
alternatively cases in which the position of the first \chandra\ source is in outstanding
agreement with the position of the \xmm/\swift\ detection, while the second \chandra\
source is right at the edge of the 3$\sigma$ uncertainty range). In these cases, we
make a judgement call ourselves and assign the unresolved detection to the resolved
source we feel is most appropriate. Where we feel unable to make a judgement, but
there are multiple ULX candidates among the resolved sources (similar to the case of
NGC\,2276 highlighted above), we we retain all of the potential resolved matches
within the master catalogue. Both of these scenarios are also indicated by the
matching flags highlighted above (again, see Table \ref{tab_matchflag}).

\begin{table}
  \caption{Definitions for the flags detailing the decision taken for any complex matches
  between the individual ULX catalogues}
\begin{center}
\begin{tabular}{c p{6.5cm}}
\hline
\hline
\\[-0.25cm]
Value & Description \\
\\[-0.25cm]
\hline
\hline
\\[-0.2cm]
NULL & No match between the catalogues \\
\\[-0.3cm]
0 & Unique match between ULX candidates \\
\\[-0.3cm]
1 & Formally more than one potential match between ULX candidates, but one is clearly preferred and assumed to be the correct match; only this match is reported \\
\\[-0.3cm]
2 & Formally more than one potential match between ULX candidates, and it is unclear which is the correct association; all potential matches are given \\
\\[-0.3cm]
3 & ULX detection consistent with several lower luminosity sources seen by the other mission in question, but their combined flux is not sufficient to explain that seen of the ULX detection, so the source is still retained \\
\\[-0.2cm]
\hline
\hline
\end{tabular}
\end{center}
\vspace*{-0.1cm}
\label{tab_matchflag}
\end{table}

\begin{figure*}
\begin{center}
\hspace*{-0.2cm}
\rotatebox{0}{
{\includegraphics[width=235pt,trim={0cm -1cm 0cm -1cm},clip]{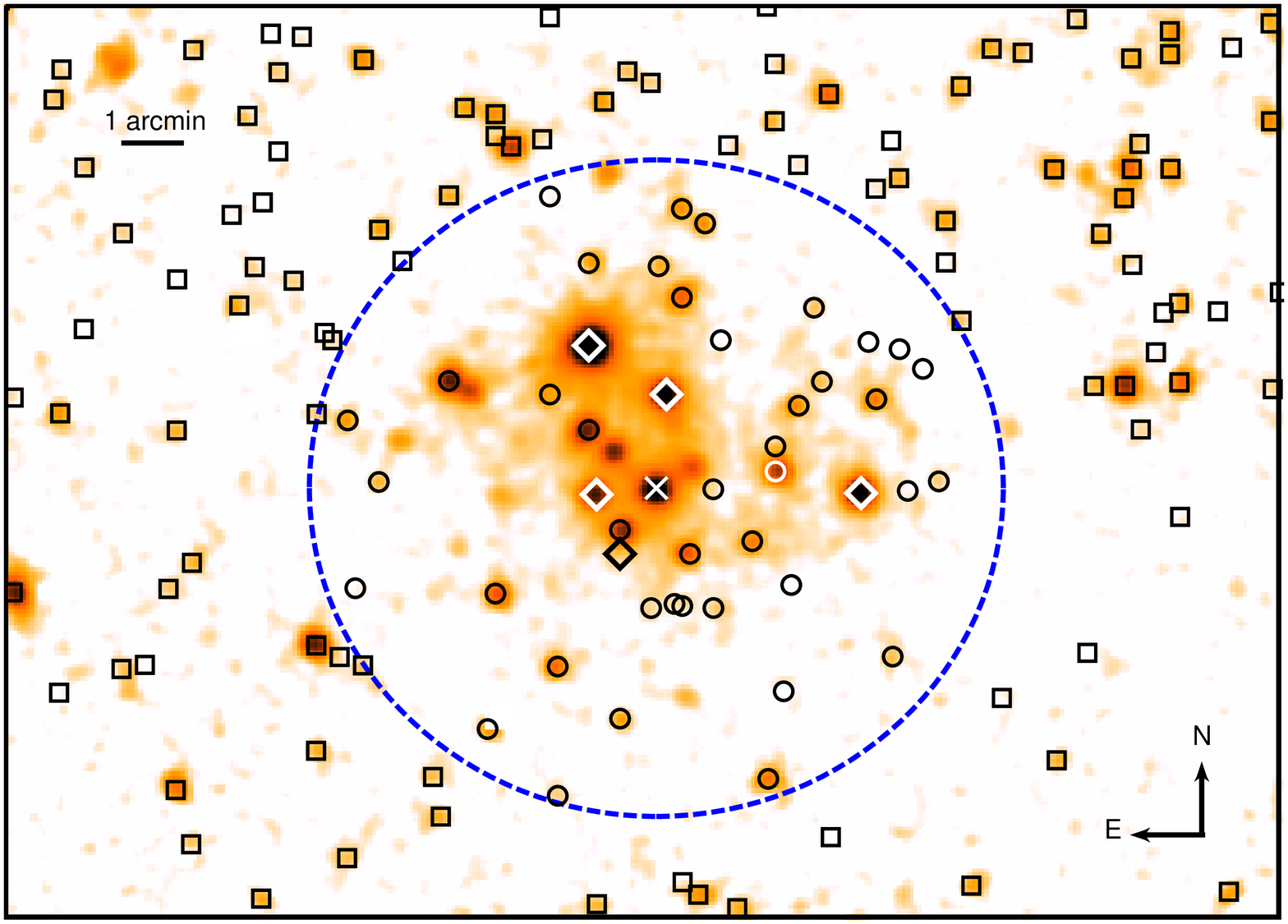}}
}
\hspace*{0.6cm}
\rotatebox{0}{
{\includegraphics[width=235pt,trim={0cm -1cm 0cm -1cm},clip]{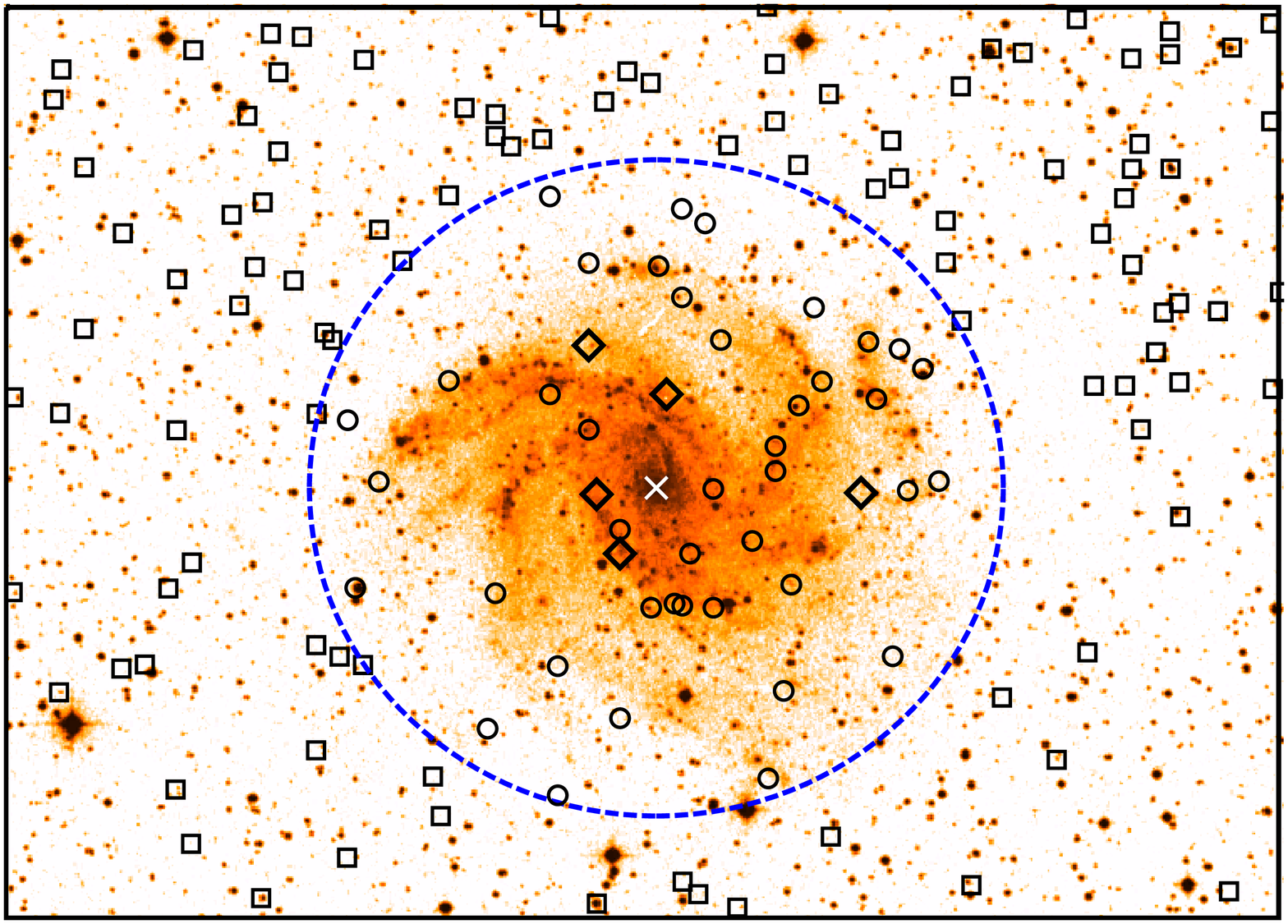}}
}
\hspace*{-0.2cm}
\rotatebox{0}{
{\includegraphics[width=235pt,trim={0cm -1cm 0cm -1cm},clip]{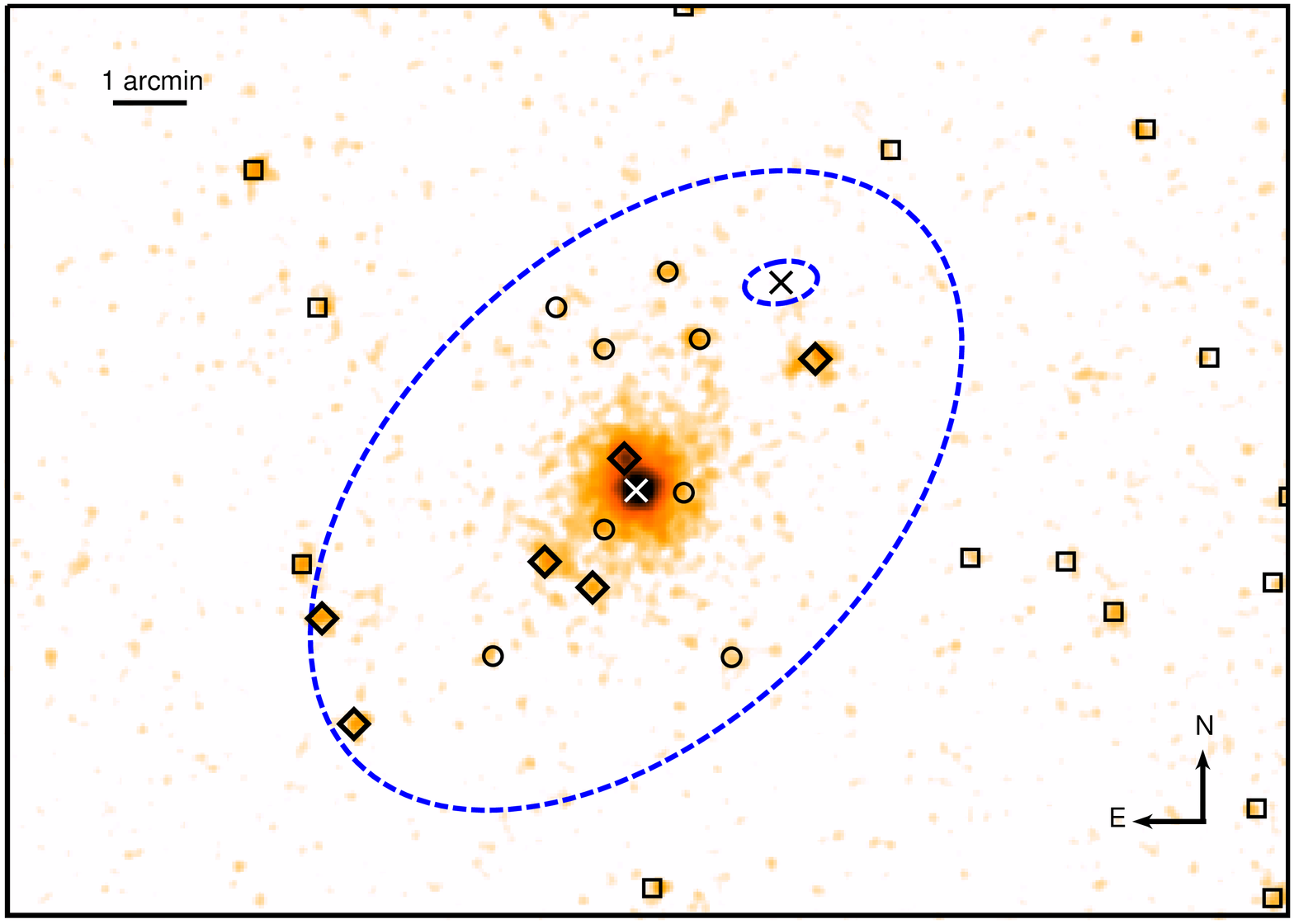}}
}
\hspace*{0.6cm}
\rotatebox{0}{
{\includegraphics[width=235pt,trim={0cm -1cm 0cm -1cm},clip]{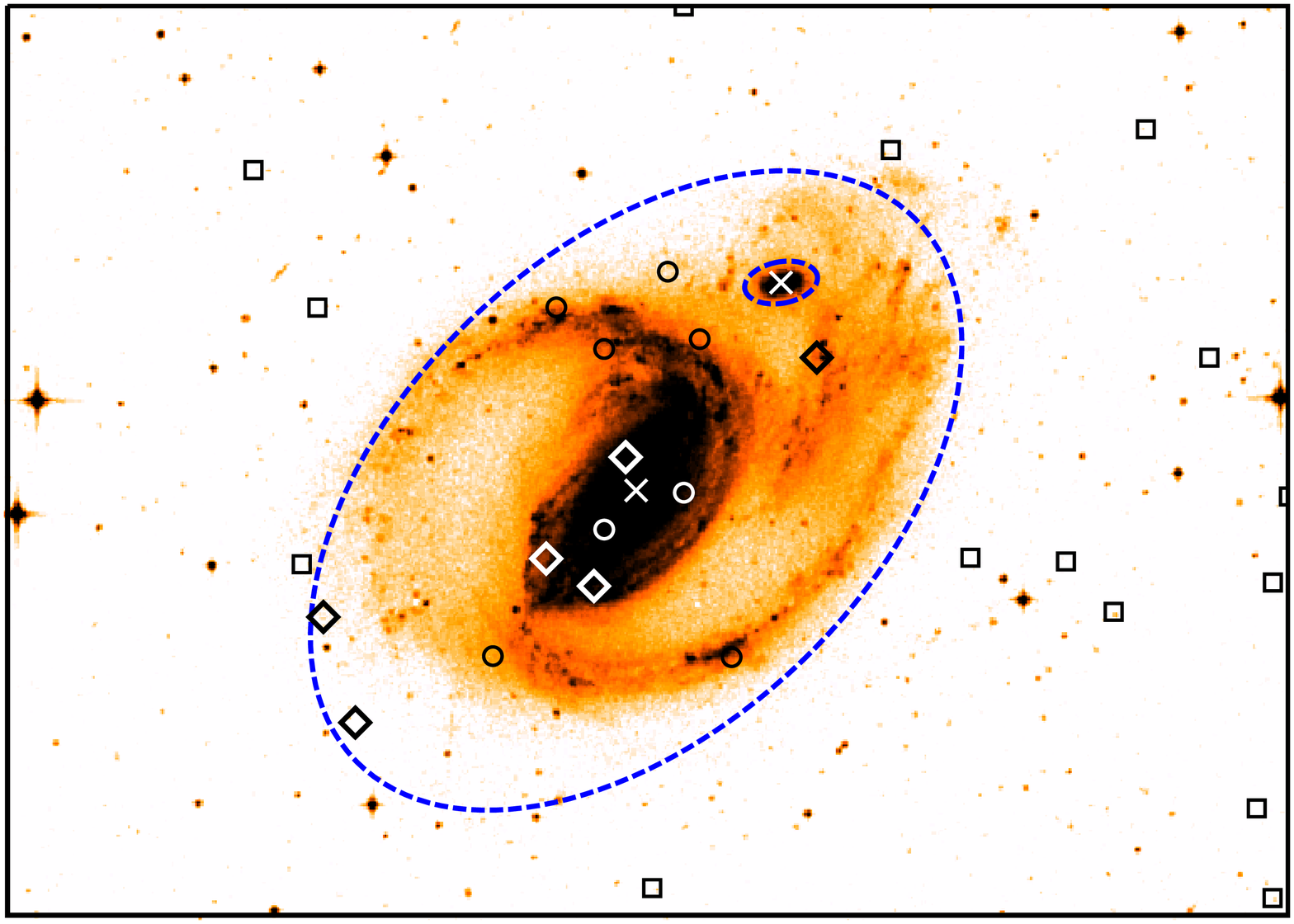}}
}
\hspace*{-0.2cm}
\rotatebox{0}{
{\includegraphics[width=235pt,trim={0cm -1cm 0cm -1cm},clip]{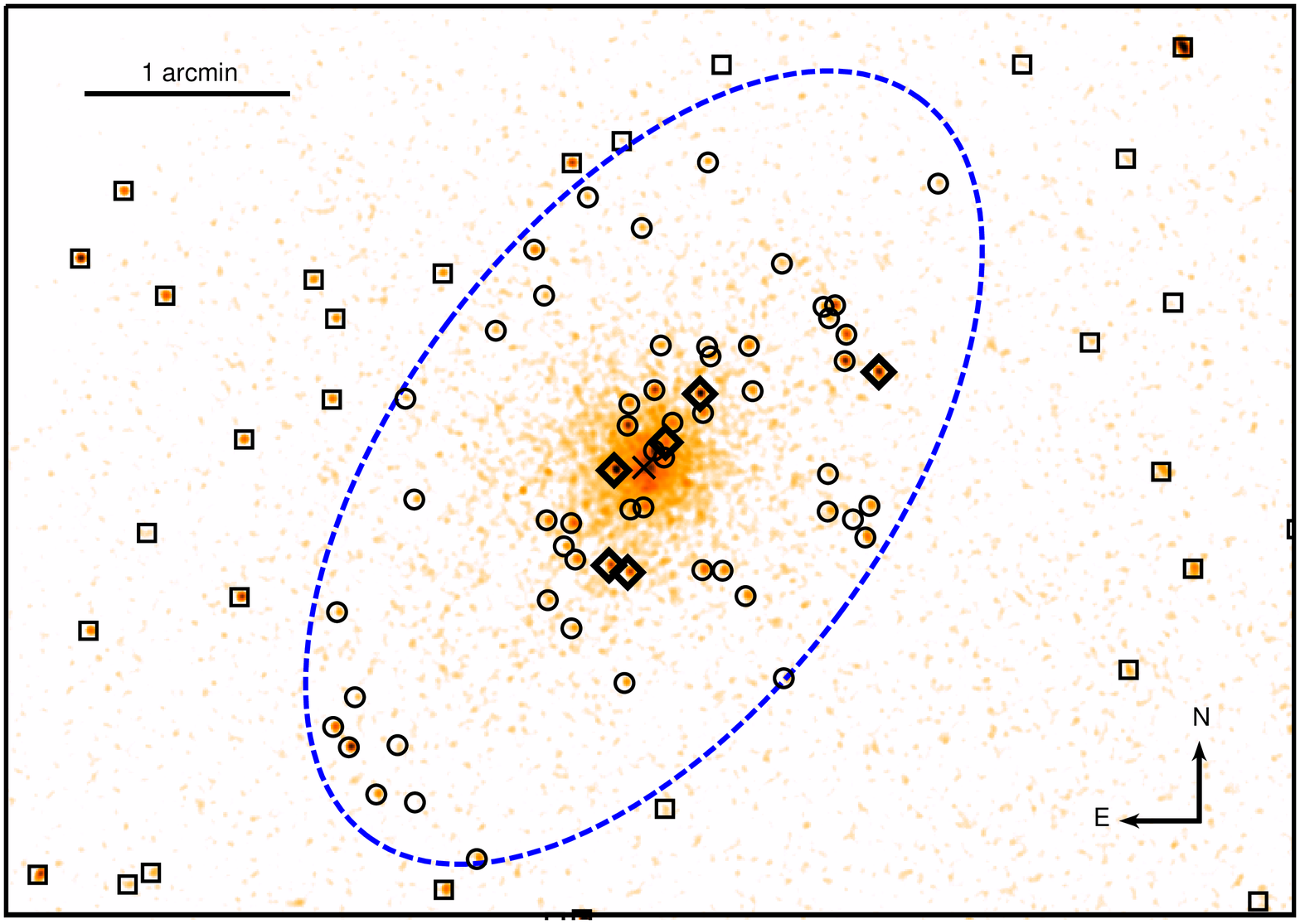}}
}
\hspace*{0.6cm}
\rotatebox{0}{
{\includegraphics[width=235pt,trim={0cm -1cm 0cm -1cm},clip]{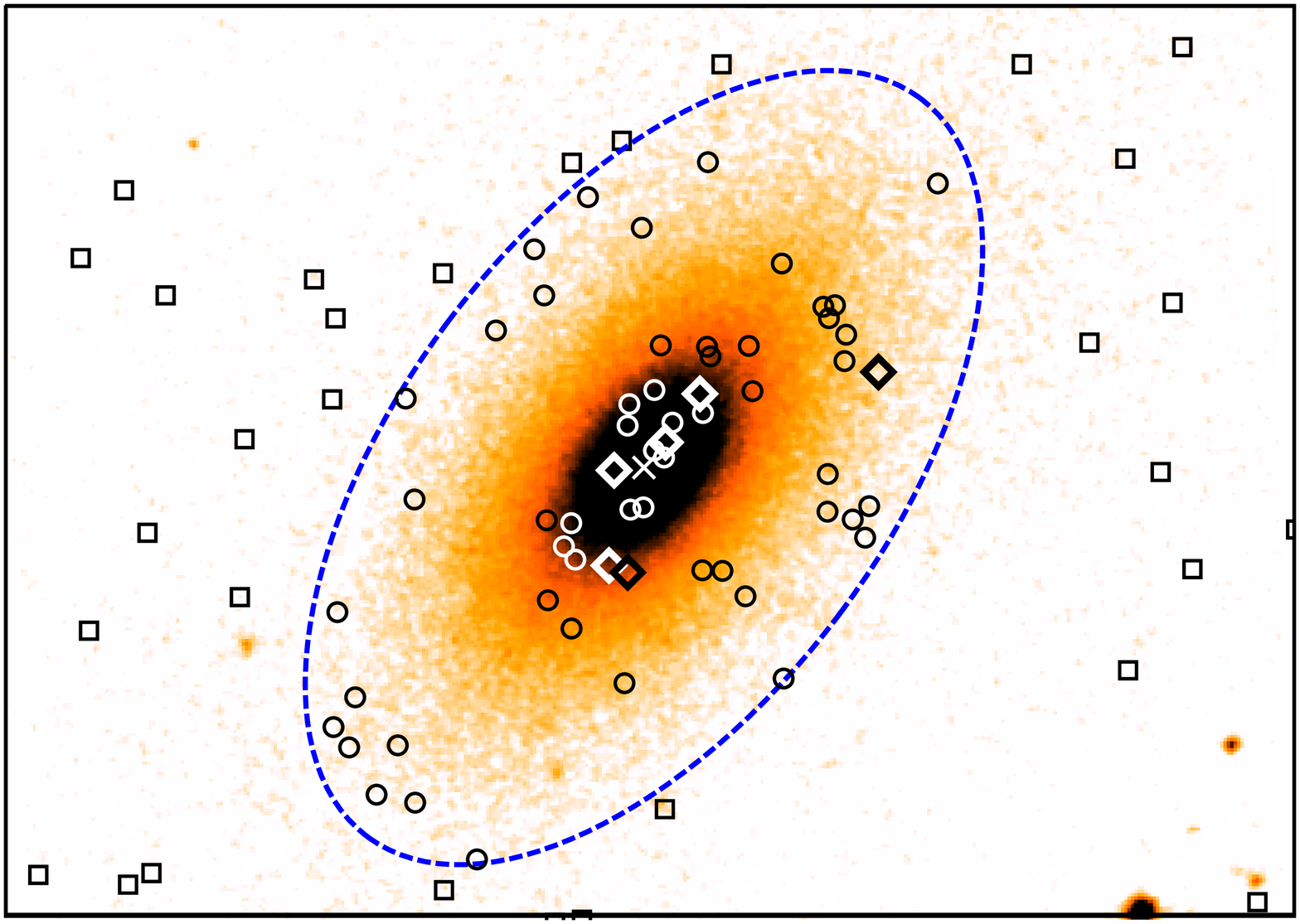}}
}
\end{center}
\vspace*{-0.3cm}
\caption{
X-ray (\textit{left}) and accompanying optical (\textit{right}) images for three example
galaxies that demonstrate some of the main stages of our source selection. For the
X-ray images, from top to bottom, we show an \xmm\ EPIC image of NGC\,6946
(OBSID 0691570101), the integrated \swift\ XRT image of NGC\,1097 (generated with
the standard online XRT pipeline; \citealt{Evans09}) and a \chandra\ ACIS image of
NGC\,720 (stack ID acisfJ0153056m134345). All are smoothed with a Gaussian of
width 3 pixels. The optical images are from the Digitized Sky Survey. In all panels, the
D25 extent of the galaxy in question is indicated with the blue dashed ellipse, and the
nuclear position with a cross (note that in the case of NGC\,1097, the small companion
galaxy NGC\,1097A is also shown). `Field' sources (\ie outside of the D25 extent) are
shown with squares, point sources within the D25 extent that do not qualify as ULX
candidates with circles, and finally sources selected as ULX candidates are shown
with diamonds, respectively; all markers are shown in either black or white simply so
that they can most easily be seen against the background images, there is no further
significance to the choice of colour. The \xmm\ and \chandra\ images represent the
deepest observation/stack of NGC\,6946 and NGC\,720 that are included in
4XMM-DR10 and CSC2, respectively. However, owing to the variable nature of the ULX
population (see \eg\ \citealt{Earnshaw19} for NGC\,6946 in particular), not all sources
identified as ULX candidates are necessarily visible in these X-ray images.
}
\label{fig_image}
\end{figure*}

Having merged the \xmm, \swift\ and \chandra\ data as best we can, we now address
the presence of one more class of known contaminants, X-ray transients associated
with one-off explosive events (\ie\ supernovae). Although certainly not all do, these
events can reach ULX luminosities, and would then be selected by our process (given
our interest in genuinely transient ULXs) even though they are clearly not
accretion-powered X-ray binaries. This is particularly relevant here given our use of
\swift\ data, since one of \swift's main focuses is rapid follow-up of transient events.
We therefore correlate our master catalogue with the positions of known supernovae
recorded in the Open Supernova Catalogue (\citealt{OpenSN}; note that this includes
both supernovae that have occurred since \xmm, \chandra\ and \swift\ have been
observing and more historic supernovae). To do so, we prioritise X-ray source
positions from \chandra, \xmm\ and \swift\ in that order (\ie\ in cases where a source
is detected by all three observatories, we use the \chandra\ position for this match),
and perform the match using search radii of 3$''$, 5$''$ and 10$''$ for \chandra,
\xmm\ and \swift\ positions, respectively. However, in order to determine whether
the X-ray source is really associated with the transient in question we also examine
the relative timing of the event and the first detection of the X-ray source (hence our
decision to only apply this filter to the merged dataset, where we can most robustly
determine when the source was first detected). X-ray sources that are positionally
coincident with supernovae, but which were detected as ULXs significantly before
the event occurred are deemed to be unrelated to the supernova and retained in our
sample. However, sources positionally coincident with known transients that have
only been detected after the event occurred are assumed to be associated with the
supernova, and so are excluded from our final sample.

We also match our remaining sample against both the NED and SIMBAD databases
in order to identify and remove any further non-ULX contaminants that have been
identified in the literature (uncatalogued AGN, stars, supernovae). We adopt the
same spatial matching procedure as for the Open Supernova Catalogue, prioritising
\chandra, \xmm\ and \swift\ positions in that order and using matching radii of 3$''$,
5$''$ and 10$''$. For any further supernovae identified, we also again consider the
date of the first X-ray detection when deciding whether the X-ray source should be
removed. We then remove any remaining sources obviously associated with the
host-galaxy AGN that have been missed by our nuclear cut (\eg\ sources with
$L_{\rm{X}} \geq 10^{42}$\,\ergps\ that lie just outside our nuclear exclusion radii or,
in the case of Centaurus A, are located in the X-ray emission from the AGN jet;
\citealt{Hardcastle07}) as well as a number of sources for which we are aware of
follow-up studies that have previously found the ULX candidate to be an
uncatalogued background quasar/foreground star (\citealt{Dadina13, Heida13,
Sutton15, Gou16}).

Finally, after all of the above steps, we find that the remaining sample contains a
number of highly clustered sources which only appear in 2SXPS and actually seem
to be associated with the bright diffuse emission known to be present in the M82
galaxy (\eg\ \citealt{Griffiths00, Lopez20}), even though 2SXPS is intended to be a
dedicated point source catalogue. This is likely related to the typical snapsnot
nature of \swift\ XRT observations; with such short exposures random Poisson
fluctuations from the diffuse emission may more easily be mistaken as point
sources. 2SXPS notes all of the potential aliases for each entry, and many of these
M82 sources are listed as potentially being aliased with each other. We therefore
also manually inspect X-ray images -- both the images integrated over the duration
of the \swift\ mission and specifically taken from the observation corresponding to
the reported best detection for the XRT, and, where available, any CSC2 \chandra\
images as well -- for all of the 2SXPS sources which have not also been identified
as a ULX candidate in either of our 4XMM-DR10 or CSC2 subsamples and are
listed as having other potential 2SXPS aliases. Any sources which we judge to be
likely associated with diffuse emission (similar to the M82 case) are removed from
the final sample. During this process, if a source is aliased with another genuine
point source (as opposed to being part of a cluster of sources associated with
extended emission), we also make a judgement over whether these are likely the
same source, and retain only one entry in these cases.

\begin{table*}
  \caption{The final sample of ULX candidates compiled from the 4XMM-DR10, 2SXPS and
  CSC2 catalogues}
\begin{center}
\begin{tabular}{l c c c c}
\hline
\hline
\\[-0.2cm]
 & 4XMM-DR10 & 2SXPS & CSC2 & Combined Sample \\
\\[-0.25cm]
\hline
\hline
\\[-0.2cm]
Number of ULX Candidates & 641 & 501 & 1031 & 1843 \\
\\[-0.3cm]
\hspace{0.5cm} (with multiple ULX detections in the parent catalogue) & 177 & 291 & 246 & 702 \\
\\[-0.3cm]
\hspace{0.5cm} (seen as a ULX by multiple observatories) & 241 & 173 & 209 & 293 \\
\\[-0.3cm]
\hspace{0.5cm} (HLX candidates) & 22 & 36 & 17 & 71 \\
\\
Host Galaxies & 403 & 269 & 548 & 951 \\
\\[-0.3cm]
\hspace{0.5cm} (average distance, Mpc) & $62.3 \pm 3.5$ & $34.8 \pm 2.7$ & $83.8 \pm 3.8$ & $74.7 \pm 2.7$ \\
\\[-0.3cm]
\hspace{0.5cm} (containing multiple ULX candidates) & 130 & 89 & 190 & 333 \\
\\[-0.2cm]
\hline
\hline
\end{tabular}
\end{center}
\vspace*{-0.1cm}
\label{tab_sample}
\end{table*}

\section{The Final Sample}
\label{sec_sample}

Our final sample of ULX candidates consists of \nulx\ individual sources residing
in \nhostgals\ host galaxies. The catalogue will be made available to the public,
and will be comprised of four tables. The first is a `master' list formatted to have
one row entry per source, summarising some key information and detailing which
combination of \xmm, \swift\ and \chandra\ have reported the source as a ULX.
We stress that we are focused only on the detections of these sources that meet
the ULX luminosity threshold here (\ie $L_{\rm{X}} \geq 10^{39}$\,\ergps); if an
\xmm\ ULX candidate does not have a \chandra\ counterpart reported, for
example, this does not necessarily mean that \chandra\ has not detected that
source, only that \chandra\ has not seen it at a flux that would correspond to the
ULX regime. The other three tables provide the full details of the 4XMM-DR10,
2SXPS and CSC2 entries for the ULX-level detections of these sources. These
follow the formats of the data used to compile these subsamples of ULX
candidates in the first place (i.e. the \xmm\ and \chandra\ tables have one row
entry per observation of a ULX candidate, while the \swift\ table just has one row
entry per ULX candidate).

Some statistics for the full sample and the individual 4XMM-DR10, 2SXPS and
CSC2 sub-samples are given in Table \ref{tab_sample}, and we show examples of
our source selection in Figure \ref{fig_image} for each of the \xmm, \swift\ and
\chandra\ observatories. By number, the CSC2 component contributes the most
sources to our final sample, followed by 4XMM-DR10 and then 2SXPS. The latter
still makes a very significant contribution though. There is obviously notable
overlap between the individual subsamples (\ie some sources are detected as
ULXs by multiple missions), as also detailed in Table \ref{tab_sample} and in the
master table provided. Of our \nulx\ individual sources, \nulxallmiss\ are
detected at ULX luminosities in all three of our contributing source catalogues.

\begin{figure}
\begin{center}
\hspace*{-0.35cm}
\rotatebox{0}{
{\includegraphics[width=235pt]{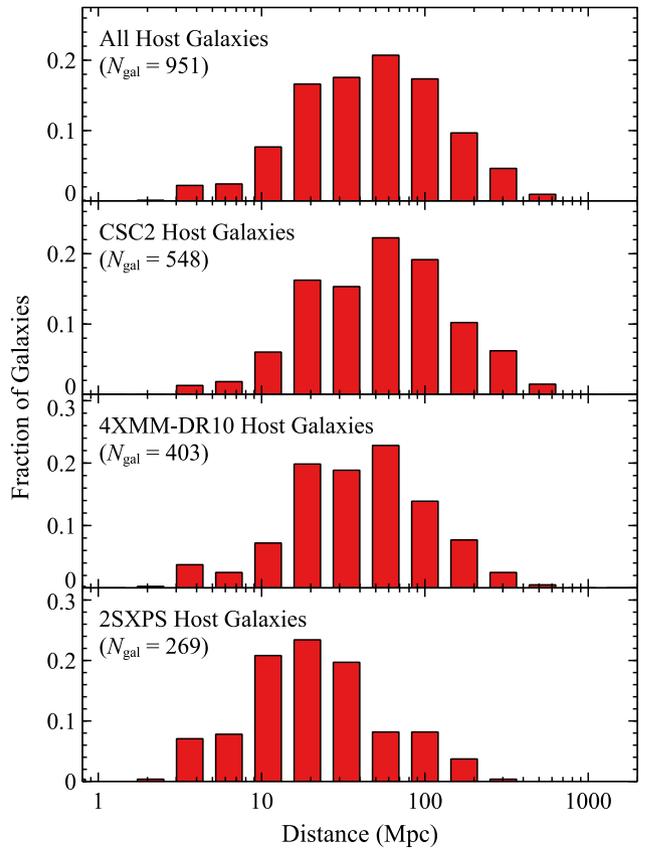}}
}
\end{center}
\vspace*{-0.3cm}
\caption{
Distance distributions for ULX host galaxies, showing the full multi-mission sample
(top), and the CSC2, 4XMM-DR10 and 2SXPS subsets, respectively (lower panels).}
\label{fig_ulx_galdist}
\end{figure}

As expected, given the known connection between ULXs and recent star formation
(\citealt{Swartz09, Mineo12, Lehmer19}), the majority of our ULX host galaxies
with morphology information available are spiral galaxies (\fracspiral\%, using the
T-type ranges defined above). We also plot the distribution of host galaxy distances
in Figure \ref{fig_ulx_galdist} for the full sample and each of the individual catalogue
subsamples. There is significant overlap in the individual distributions, but typical
host galaxy distances are lowest for the 2SXPS subsample, and largest for the
CSC2 subsample, as the latter has the best sensitivity to faint point sources among
the X-ray catalogues considered owing to both the low background and superior
imaging capabilities of \chandra. This allows \chandra\ to more easily detect ULX
candidates out to larger distances than either \xmm\ or \swift, and thus the CSC2
subsample ends up making the largest contribution to our final sample.

Of our \nhostgals\ host galaxies, \nhostgalsmult\ are found to host multiple
ULX candidates. As our primary interest is focused on individual sources, and
our sample selection is highly non-uniform, we do not make any attempt to
correct for (in)completeness in any galaxies observed with insufficient depth to
reach luminosities of 10$^{39}$\,\ergps, so this should likely be considered a
lower limit for ULX multiplicity in these hosts. The most extreme example is
NGC\,2207 -- one half of an interacting galaxy pair (the other being IC\,2163;
\citealt{Eskridge02}) -- which appears to host an astonishing 34 ULX
candidates, the majority of which (31) are contributed by the CSC2 catalogue.
This is notably larger than the 21 ULXs reported to reside in NGC\,2207/IC\,2163
by \cite{Mineo13}, likely due to additional \chandra\ observations being included
in CSC2 and our explicit consideration of long-term variability in selecting our
ULX sample. Owing to the interacting nature of these galaxies, it is not surprising
that there should be a large number of ULXs. It is nevertheless worth noting that
there seems to be some disagreement over the distance to NGC\,2207 in the
literature. The distance we have adopted here is $D = 36.4$\,Mpc, which is
based on the recession velocity reported in HyperLEDA. This distance is very
similar to that reported based on the supernova SN1975A which occurred in
NGC\,2207 ($D = 39.6$\,Mpc; \citealt{Arnett82}), which is adopted by
\cite{Mineo13}. However, the more recent estimates from the Tully-Fisher
method typically seem to imply a distance of $D \sim 17$\,Mpc
(\citealt{Russell02, Theureau07}). Should this be correct, only 7 of our sources
in NGC\,2207 would still be considered ULXs. However, our assumption is that
the supernova-based distance is the most reliable here, and so our luminosity
estimates should be reasonable.

We also note that among the \nulx\ ULX candidates, our catalogue contains \nhlx\
`hyperluminous' X-ray source (HLX) candidates.\footnote{Note that for in order for
sources detected in 2SXPS to be considered good HLX candidates we apply a
similar criterion to our initial source selection procedure, such that if the peak
luminosity is not well constrained (average fractional uncertainty of $>$40\%) then
the source has to either have a better-constrained average luminosity that is also in
the HLX regime, or at least two separate XRT observations that place it in the HLX
regime.} These are the most extreme members of the ULX population, exhibiting
luminosities of $L_{\rm{X}} \geq 10^{41}$\,\ergps. Owing to their astonishing
luminosities, such sources are often considered the best candidates for IMBH
accretors. Indeed, two of the sources discussed as the leading IMBH candidates in
the literature, M82 X-1 and ESO\,243--49 ULX1, are found among this population.
However, it is also worth noting that one of the known ULX pulsars, NGC\,5907
ULX1, also reaches luminosities of $L_{\rm{X}} \sim 10^{41}$\,\ergps\ (\citealt{Israel17,
Fuerst17ngc5907}), despite being powered by a neutron star. Nevertheless, these
sources are still of particular interest, and our new HLX candidates will be discussed
in more detail in future work (A. D. Mackenzie, \textit{in prep.}).

\subsection{Comparison with Other ULX Catalogues}
\label{sec_prev_ulxcat}

The first major effort to search for ULXs among any of the X-ray source catalogues
considered here was presented by \cite{Kovlakas20}, who also searched CSC2 for
ULX candidates. Although both the approach taken and the input galaxy sample
used are quite similar in both cases, there are also a couple of notable differences.
First and foremost, we have considered the \chandra\ data down to the
observation-by-observation level, in order to select sources based on their peak
flux and specifically include transient ULXs in our sample, while \cite{Kovlakas20}
base their luminosity selection on the flux recorded in the longest uninterrupted
segment of \chandra\ data (which is not necessarily the peak flux exhibited by the
source). Second, we have taken a much more conservative approach to excluding
potential nuclear sources associated with our host galaxies. \cite{Kovlakas20} flag
a source as `nuclear' if it is within 3$''$ of the nominal galaxy centre, while we both
consider the position error on the X-ray detection and utilize a much larger
minimum exclusion radius (6.1$''$). This is based on our empirical assessment of
the separation between the nominal centre of the host galaxies and sources that
we consider likely to be their AGN (those that appear to have $L_{\rm{X}} \geq
10^{42}$\,\ergps). Our more conservative approach does mean that our catalogue
is likely cleaner with regards to any remaining contamination from AGN in our host
galaxies, but this will come at the cost of excluding a larger number of \textit{bona
fide} ULXs from our sample, particularly given that the spatial density of ULXs is
seen to increase towards the  galaxy centres (\citealt{Swartz11, Wang16,
Kovlakas20}). Nevertheless, this is a more appropriate approach given that our
primary motivation is to find individual sources that are of interest for follow-up
studies; detailed follow-up of ULXs within a few arcseconds of the nuclear
position will not realistically be feasible for the majority of our current and planned
 X-ray facilities if the nuclear black hole is even reasonably active. Despite these
differences, though, there is naturally a fairly significant degree of overlap (754
sources) between our sample and sources that would qualify as ULXs in
\cite{Kovlakas20}.

More recently, \cite{Inoue21} have also searched for CSC2 for ULX candidates.
However, a major difference between these works is that \cite{Inoue21} have
used a much smaller catalogue of input galaxies than that used here, derived by
combining IRAS galaxies with the CNG catalogue. Furthermore, while they do
consider the observation-by-observation data provided in CSC2, they use the
flux from the longest individual observation when computing luminosities, which
again is not necessarily the peak flux exhibited by the sources in question (which
is what we are interested in here), and we have again been more conservative in
our treatment of nuclear sources. Although their work primarily focuses on
CSC2, the final catalogue does also include sources selected from 4XMM-DR9
and 2SXPS, and so is therefore conceptually similar to our multi-mission
approach. There is not a lot of specific detail provided for these latter analyses,
unfortunately, but the approach is stated to be broadly similar to that used for
CSC2, and so similar differences between the two works are presumably present
here too. In addition, another key difference with regard to the 2SXPS analysis is
that they appear to have only made use of the average fluxes from \swift, while
we have considered the peak flux (where this is considered reliable). Furthermore,
we have used an even more recent release of the 4XMM survey here.
Nevertheless, despite these differences, there is again some notable overlap of
357 sources in total (251, 107 and 149 from \chandra, \xmm\ and \swift,
respectively).

Finally, \cite{Barrows19} also utilize CSC2, but only search specifically for HLXs
within the SDSS-DR7 galaxy sample. However, the spatial offsets relative to the
central galaxy positions would result in the majority of their HLX candidates being
considered as nuclear sources with our more empirical approach to this stage
of the catalogue production. Only one of our sources is also present in the
\cite{Barrows19} catalogue, 2CXO\,J155910.3$+$204619, and we have assigned
this to a different (and closer) host galaxy, giving it a much more modest
luminosity of $L_{\rm{X}} \sim 2.5 \times 10^{39}$\,\ergps. As such, our sample
of HLX candidates differs entirely to that presented by \cite{Barrows19}.

In addition to these more recent works, we also match our new catalogue against
a series of other archival ULX catalogues, which have been derived using
previous generations of X-ray surveys. In these cases we match by position, as
they are not drawn from any of the exact X-ray catalogues used here (and thus
do not have identical naming conventions). Similar to our final matching against
the NED and SIMBAD for remaining contaminants, we split our ULX catalogue
into sources where the best position comes from \chandra, from \xmm\ and from
\swift, and then individually match these sub-sections against each of the
archival ULX catalogues in turn. The matching radius used for each comparison
depends on the origins of the data being compared, and always corresponds to
the larger of the typical positional uncertainties associated with the two input
tables. As before, positions from \chandra, \xmm, \swift\ and \rosat\ are
considered to have typical uncertainties of 3$''$, 5$''$ and 10$''$, respectively,
and \rosat\ positions are considered to have a typical uncertainty of 20$''$. For
example, when comparing the subset of our catalogue with \chandra\ positions
against another catalogue derived from \chandra\ data, we would use a
matching radius of 3$''$, but comparing the same subset against a catalogue
derived from \rosat\ observations, we would use a matching radius of 20$''$
instead. For this analysis, we simply note all potential matches. The catalogues
we match against are listed in Table \ref{tab_oldcat}.

Based on all of these matches, we find that \nulxnew\ of the ULX candidates
presented here are completely new, \ie\ do not seem to appear in any of the
other ULX catalogues considered, and \nulxrecent\ have only recently been
catalogued as a ULX, \ie they only appear in catalogues based on the latest
generation of X-ray source catalogues (this work, \citealt{Barrows19},
\citealt{Kovlakas20} and \citealt{Inoue21}). Of these `new' and `recent' ULX
candidates, \nhlxnew\ and \nhlxrecent, respectively, are HLX candidates. We
stress that even if a source is considered `new' in this respect, this does not
necessarily mean the sources are completely unknown, only that it has not
been formally catalogued as a ULX previously. For example, NGC\,300 ULX1 is
considered `new' here, even though this source is one of the few known ULX
pulsars (\citealt{Carpano18}), and as such has received significant individual
attention (\citealt{Walton18crsf, Kosec18, Vasilopoulos18, Vasilopoulos19,
Heida19}).

\begin{table}
  \caption{Details of the archival ULX catalogues against which our new archive is compared}
\begin{center}
\begin{tabular}{c p{5cm}}
\hline
\hline
\\[-0.25cm]
Catalogue & Primary Source \& Notes \\
\\[-0.25cm]
\hline
\hline
\\[-0.2cm]
\cite{Colbert02} & \rosat\ \\
\\[-0.3cm]
\cite{Swartz04} & \chandra\ \\
\\[-0.3cm]
\cite{LiuBregman05} & \rosat\ \\
\\[-0.3cm]
\cite{LiuMirabel05} & Literature (incl. \rosat, so positions treated as having \rosat\ accuracy) \\
\\[-0.3cm]
\cite{Liu11} & \chandra\ \\
\\[-0.3cm]
\cite{Swartz11} & \chandra\ \\
\\[-0.3cm]
\cite{WaltonULXCat} & \xmm\ (specifically 2XMM) \\
\\[-0.3cm]
\cite{Gong16} & \chandra\ (only $L_{\rm{X}} \geq 3 \times 10^{40}$\,\ergps) \\
\\[-0.3cm]
\cite{EarnshawULXcat} & \xmm\ (specifically 3XMM-DR4) \\
\\[-0.3cm]
\cite{Barrows19} & \chandra\ (specifically CSC2 HLXs) \\
\\[-0.3cm]
\cite{Kovlakas20} & \chandra\ (specifically CSC2) \\
\\[-0.3cm]
\cite{Inoue21} & Mainly  \chandra\ (specifically CSC2), but also includes \xmm\ and \swift\ (specifically 4XMM-DR9 and 2SXPS) \\
\\[-0.2cm]
\hline
\hline
\end{tabular}
\end{center}
\vspace*{-0.1cm}
\label{tab_oldcat}
\end{table}

\subsection{Estimation of Unknown Contaminants}
\label{sec_contam}

Although we have taken significant measures to try and remove known contaminants,
these processes can never be perfect, and so we stress that there will still be a
significant contribution of sources that are not actually ULXs in our final sample of ULX
candidates. By far the majority of these will be foreground/background sources that
coincidentally appear to be associated with the host galaxies in question in projection,
but have just not been formally identified/catalogued as such in the databases we have
utilized (and thus have not been removed by our effort to identify and exclude these
sources). Although we cannot remove these sources, it is still important to quantify
their likely contribution.

In order to do so, we broadly follow the approach taken in \cite{WaltonULXCat}, and
subsequently \cite{Sutton12} and \cite{EarnshawULXcat}. This involves a calculation of
the total expected number of sources that would be resolved from the cosmic X-ray
background (CXB) given our selection criterion, the sensitivity of the observations from
which the 4XMM, 2SXPS and CSC2 X-ray catalogues have been generated, and
the full set of galaxies in our catalogue that have been covered by the observations
that contribute to these X-ray catalogues (not just those galaxies that have ULX
detections). These estimates are then compared to the number of sources remaining in
our catalogue, after accounting for the number of identified foreground/background
sources that have already been filtered out, in order to estimate the remaining fractional
contribution from these contaminants. 

In order to estimate the total expected number of contaminants, we make use of
empirically determined forms of the \NS\ curves which quantify the number of sources
per square degree ($N$) resolved from the CXB as a function of flux sensitivity ($S$).
These are combined with observational sensitivity maps in order to estimate the
number of background sources each galaxy in our input sample that has been
observed should contribute. Sensitivity maps for the observations from which the
4XMM and CSC2 catalogues are compiled are provided as part of these data releases,
but are not available for the 2SXPS catalogue at the time of writing. We therefore
focus our calculations on the 4XMM and CSC2 data, performing this calculation for
each dataset separately; as will be clear below, the expected level of contamination
for these datasets are very similar, and so we still expect these results to hold overall.

For CSC2, since the initial source detection is performed using `stacks' of observations
(a stack is defined as a group of observations for which the aimpoints are all within 1
arcminute; see the CSC2 documentation), we use the sensitivity maps generated for
these stacks in our analysis. These are provided for all of the energy bands considered
in the CSC2 catalogue. However, as noted by \cite{WaltonULXCat}, owing to absorption
in the apparent host galaxies (which lie between us and any background AGN) these
calculations are most robust at higher energies, and so we limit ourselves to the hard
band (2--7\,keV) ACIS maps that correspond to the `true' detection threshold
to match our data selection (the HRC makes a negligible overall contribution here). We
also make use of the \NS\ curves recently published by \cite{Masini20}, who present an
expression for the same 2--7\,keV band.

There are two limiting fluxes to consider here. The first is set by our selection of sources
that appear to have $L_{\rm{X}} \geq 10^{39}$\,\ergps. For each galaxy we work out the
hard band flux that would correspond to a broadband luminosity of $10^{39}$\,\ergps,
$S_{\rm{ulx}}$, based on the distance to the galaxy and the fraction of the broadband
flux that would appear in the hard band. We use a coarse representation of the average
spectral shape for ULXs below 10\,keV (\eg\ \citealt{Stobbart06, Gladstone09,
Pintore17}): an absorbed powerlaw spectrum with $\langle N_{\rm{H}} \rangle = 3 \times
10^{21}$\,\pcmsq\ and $\langle \Gamma \rangle = 2.1$. The second flux is the limiting
sensitivity of the stack in question, $S_{\rm{obs}}$, provided by the sensitivity maps.
The relevant limiting sensitivity for use with the \NS\ curve is then the larger of these
two values, such that if an observation is sensitive enough to detect sources at lower
luminosities, these are not included in our estimated number of contaminants. For each
of the galaxies covered by CSC2 we use the \NS\ curve and the appropriate limiting
sensitivity to compute maps of the number of expected background sources per pixel,
and integrate these over the area of the galaxy covered by each relevant \chandra\
stack (excluding the typical area excised around the central galaxy location by our
nuclear cut). For each of the galaxies covered by CSC2 data (again, not just those with
ULX detections), we perform this calculation for every available stack. We then select
the stack that would give the largest number of contaminants, and sum these values
over all of the galaxies covered by CSC2 stacks to compute the total number of
expected contaminants prior to the removal of any known foreground/background
sources. From this, we compute the remaining fractional contamination among the
CSC2 ULX candidates by comparing the expected number of remaining contaminants
to the number of ULX candidates detected in the hard band for self-consistency (\ie
excluding sources that only have upper limits).

For 4XMM, the sensitivity maps are only provided for the full band (0.2--12.0\,keV)
and are based on the combined sensitivity for all of the EPIC detectors (see Section 9
in \citealt{4XMM}). However, as stated above, it is preferable to work in the hard band
here. Furthermore, suitable \NS\ curves are not currently available for the full \xmm\
bandpass; aside from work focusing specifically on \chandra, \NS\ curves are
determined almost exclusively for the 0.5--2.0 and 2--10\,keV bands. It is therefore
necessary to correct the results from the available broadband maps to one of these
bands, and we again choose the harder band, but this is not a trivial process. To do
so, we also make use of the hard band (2--12\,keV) sensitivity maps computed as
part of the \cite{EarnshawULXcat} ULX catalogue for the majority of observations
that make up 4XMM-DR10 (specifically those that make up 3XMM-DR4). However,
these are computed using a different approach (see \citealt{Carrera07} and
\citealt{Mateos08}), consider each of the EPIC detectors separately, and adopt a
different detection threshold (the hard band maps are computed for a
$\sim$4$\sigma$ detection in a single detector, while the broadband maps are
computed for a $\sim$3$\sigma$ detection combining all the EPIC detectors), further
complicating the situation. 

For each galaxy covered by these earlier observations, we therefore process the
available hard band sensitivity maps for each of the detectors in a similar manner as
above, using the appropriate \NS\ curve published by \cite{Cappelluti09} but only
considering for the limiting observational sensitivity ($S_{\rm{obs}}$) for the time
being, and note the results for the detector that predicts the largest number of
contaminants. We also process the broadband maps for the same observations by
computing the fraction of the broadband flux in both the softer (0.5-2.0\,keV) and
harder (2--10\,keV) bands using the spectral form assumed in their generation (an
absorbed powerlaw with $N_{\rm{H}} = 1.7 \times 10^{20}$\,\pcmsq\ and $\Gamma
= 1.42$, typical for CXB sources), processing these updated maps in turn using the
relevant \NS\ curves, again only considering $S_{\rm{obs}}$, and averaging the
final results to obtain an estimate for the number of contaminating sources the
broadband maps would imply. For each galaxy covered by these earlier
observations, we then compare the results from the broadband and the hard band
maps to compute an empirical correction for the former; we find this correction to
be a factor of 9. We then process the full set of 4XMM-DR10 broadband sensitivity
maps using this correction to produce maps of the expected number of hard band
CXB sources. At this point, we also consider the number of contaminants implied
by the second limiting sensitivity, $S_{\rm{ulx}}$, and update the maps accordingly.
Similar to before, we then integrate these maps over the galaxy area covered by
every observation of that galaxy (again excluding the typical area around the
central position excised by our nuclear cut, and again considering all galaxies
covered by 4XMM-DR10) and pick the observation that gives the largest number of
hard-band contaminants for each galaxy. We then sum these values to compute the
total number of expected contaminants prior to the removal of any known
foreground/background sources, and finally compute the remaining fractional
contamination among the 4XMM ULX candidates (comparing the expected number
of remaining contaminants against the number of ULX candidates that are detected
at the 4$\sigma$ level in any of the EPIC detectors for self-consistency).

Based on these approaches, and the numbers of known foreground/background
contaminants already removed, we estimate fractional contaminations of $(23 \pm
2)$\%, and $(18 \pm 3)$\% for our CSC2 and 4XMM-DR10 ULX candidates,
respectively (quoted uncertainties are due to counting statistics, and are
1$\sigma$). These values are sufficiently similar that, even though the relevant
sensitivity maps are not yet available for 2SXPS, we still expect that an overall
fractional contamination of $\sim$20\% is relevant for the whole catalogue.

\section{NGC\,3044 ULX1 -- A New Extreme ULX}
\label{sec_ulxs}

\begin{figure}
\begin{center}
\hspace*{-0.2cm}
\rotatebox{0}{
{\includegraphics[width=235pt]{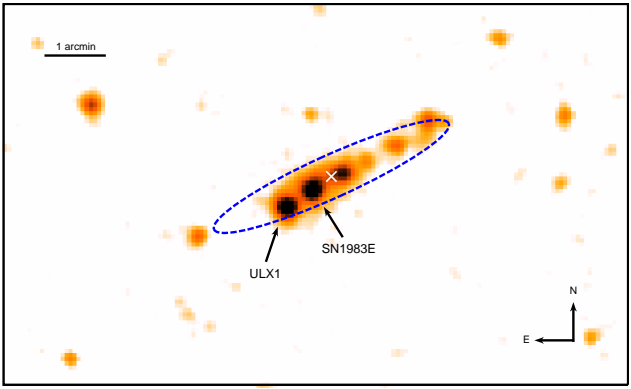}}
}
\end{center}
\vspace*{-0.3cm}
\caption{
\xmm\ image of NGC\,3044 from OBSID 0782650101. As with Figure \ref{fig_image},
the D25 extent of NGC\,3044 is shown with the blue dashed ellipse, and the
nuclear position with a white cross. The positions of the brightest ULX candidate in
NGC\,3044, which we refer to as ULX1, and the nearby supernova 1983E are
indicated.}
\label{fig_ngc3044_img}
\end{figure}

The non-uniform selection means the full ULX sample presented here is not necessarily
well suited for detailed statistical studies of the ULX population similar to those
presented by \cite{Kovlakas20} and \cite{Inoue21}. Indeed, our intention in compiling
this sample is to facilitate follow-up studies of interesting individual sources. As such,
in order to highlight the potential of our catalogue, we instead present a case study of
a new extreme ULX candidate with $L_{\rm{X,peak}} \sim 10^{40}$\,\ergps\ in the
edge-on spiral galaxy NGC\,3044 by both \swift\ and \xmm\ (see Figure
\ref{fig_ngc3044_img}). Although we have found several new HLX candidates in our
analysis, we highlight this new source in particular both because of its luminosity is still
very extreme, but also because it already has high signal-to-noise (S/N) \xmm\ data
(several thousand counts) available in the archive; as noted above, our new HLX
candidates will instead be studied in future work (A. D. Mackenzie, \textit{in prep.}).
Hereafter we refer to this source as NGC\,3044 ULX1 for simplicity, as it is the
brightest ULX candidate in NGC\,3044, but its catalogued 4XMM-DR10 and 2SXPS
IAU names are 4XMM\,J095343.8+013416 and 2SXPS\,J095343.7$+$013417,
respectively. Throughout this analysis, we assume a distance of $D = 20.6$\,Mpc to
NGC\,3044 (\citealt{Tully16}).

\subsection{Observations and Data Reduction}
\label{sec_ngc3044_obs}

NGC\,3044 has been observed on four occasions by \xmm, and on five occasions by
\swift. A log of these observations is given in Table \ref{tab_ngc3044_obs}. We primarily
focus on the \xmm\ observations here, but also process the \swift\ observations to
provide further information on the long-term variability.

The \xmm\ data for each observation are reduced following standard procedures using
the \xmm\ Science Analysis System (\sas\ v19.1.0). All of the \xmm\ observations were
taken in full frame mode. Raw observation files for the \epicpn\ and \epicmos\ detectors
are cleaned using \epchain\ and \emchain, respectively. In order to facilitate pulsation
searches, the cleaned \epicpn\ event files are corrected to the solar barycentre using the
DE200 solar ephemeris, as this has the best time resolution of the \xmm\ detectors
(73.4\,ms in full frame mode). Source products are extracted from the cleaned event
files with \xmmselect. Given the relative proximity of supernova SN1983E (separated by
$\sim$35$''$; see Figure \ref{fig_ngc3044_img}), we use circular source regions of
radius 15--20$''$, with the larger region size used for the higher flux observations (see
below). Background is estimated from a larger region of blank sky on the same detector
as ULX1. All of the observations suffer from periods of enhanced background to some
degree, and for each observation we determine the background threshold that
maximises the source S/N over the full \xmm\ band considered in our more detailed
analysis (0.3--10.0\,keV) following the method outlined in \cite{Picon04}. Only single
and double patterned events are considered for \epicpn\ ({\small PATTERN}\,$\leq$\,4)
and single to quadruple patterned events for \epicmos\ ({\small PATTERN}\,$\leq$\,12),
as recommended, and all of the necessary instrumental response files were generated
using \arfgen\ and \rmfgen. After performing the reduction separately for the two
\epicmos\ units, we combine their individual spectra using \addascaspec.

\begin{table}
  \caption{Details of the X-ray observations of NGC\,3044 ULX1 considered in this work.}
\begin{center}
\begin{tabular}{c c c c}
\hline
\hline
\\[-0.25cm]
Mission & OBSID & Start Date & Good \\
& & & Exposure (ks)\tmark[a] \\
\\[-0.3cm]
\hline
\hline
\\[-0.15cm]
\xmm\ & 0070940101 & 2001-11-24 & 6/8 \\
\\[-0.25cm]
\xmm\ & 0070940401 & 2002-05-10 & 9/21 \\
\\[-0.25cm]
\xmm\ & 0720252401 & 2013-05-06 & 9/11\tmark[b] \\
\\[-0.25cm]
\swift\ & 00092188001 & 2015-04-17 & 1 \\
\\[-0.25cm]
\swift\ & 00092188002 & 2015-04-19 & 1 \\
\\[-0.25cm]
\swift\ & 00092188003 & 2015-06-23 & 2 \\
\\[-0.25cm]
\swift\ & 00092188004 & 2015-06-24 & 2 \\
\\[-0.25cm]
\swift\ & 00092188005 & 2015-06-25 & 2 \\
\\[-0.25cm]
\xmm\ & 0782650101 & 2016-12-07 & 80/93 \\
\\[-0.2cm]
\hline
\hline
\\[-0.15cm]
\end{tabular}
\\
\flushleft
$^{a}$ \xmm\ exposures are listed for the \epicpn/MOS detectors, after
filtering for background flaring (see Section \ref{sec_ngc3044_obs}). \\
$^{b}$ ULX1 falls on a dead chip for the MOS1 detector in this observation.
\vspace*{0.3cm}
\label{tab_ngc3044_obs}
\end{center}
\end{table}

\subsection{Spectral Analysis}
\label{sec_spec}

\begin{figure}
\begin{center}
\hspace*{-0.2cm}
\rotatebox{0}{
{\includegraphics[width=235pt]{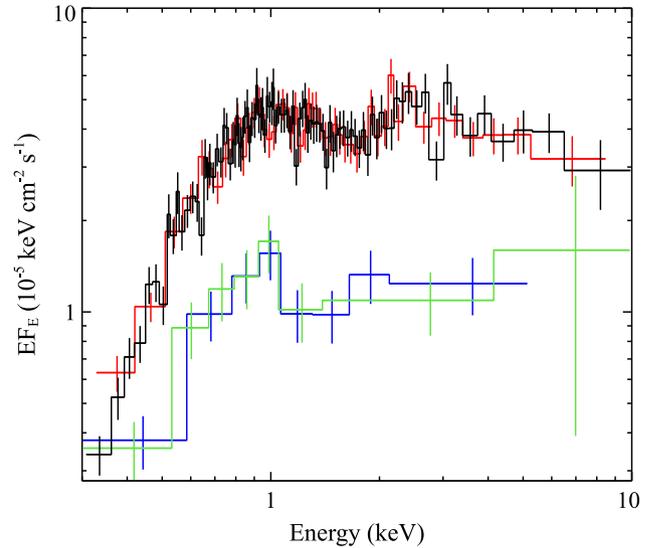}}
}
\end{center}
\vspace*{-0.3cm}
\caption{
The \xmm\ spectra of NGC\,3044 ULX1 from the high- and low-flux states seen in
the available data, unfolded through a model that is constant with energy. The \epicpn\
data are shown in black and green, and the \epicmos\ data are shown in red and blue.
The data have been rebinned for visual clarity.}
\label{fig_ngc3044_spec}
\end{figure}

Based on the 4XMM-DR10 data, the first two \xmm\ observations (2001 and 2002) both
caught NGC\,3044 ULX1 in a lower flux state, while the latter two (2013 and 2016)
caught the source in a higher flux state. We therefore combine the \xmm\ spectra from
these pairs of observations using \addascaspec\ to provide the highest S/N data possible
for these two flux regimes. These spectra are shown in Figure \ref{fig_ngc3044_spec}.

We initially begin by fitting the high-flux data with a simple absorbed powerlaw model.
We use \xspec\ for our spectral analysis (\citealt{xspec}), and allow for both the
Galactic absorption column of $N_{\rm{H,Gal}} = 2.33 \times 10^{20}$\,\pcmsq\
(\citealt{NH2016}) and further absorption at the redshift of NGC\,3044 ($z = 0.00430$)
that is free to vary in all our models. Both absorbers are modelled using \tbabs, and we
adopt solar abundances from \cite{tbabs} and absorption cross-sections from
\cite{Verner96}. We also allow for cross-normalisation constants to float between the
data from the pn and MOS detectors to account for residual calibration differences;
these factors are always within a few per cent of unity. Finally, the higher flux data are
grouped to a minimum of 25 counts per bin to facilitate the use of \chisq\ minimisation.
The absorbed powerlaw model returns a fairly steep continuum, with $N_{\rm{H,high}}
= (2.1 \pm 0.2) \times 10^{21}$\,\pcmsq\ and $\Gamma_{\rm{high}} = 2.41 \pm 0.07$
(uncertainties on the spectral parameters are quoted at the 90\% level). Unsurprisingly,
the lower flux data have a much lower S/N (in addition to the lower flux, these data have
a much lower combined exposure). We therefore group these data to just 1 count per
bin, and fit them with the same model using the Cash statistic (\citealt{cstat}). Here we
find $N_{\rm{H,low}} = 1.1^{+0.8}_{-0.7} \times 10^{21}$\,\pcmsq\ and
$\Gamma_{\rm{low}} = 2.3 \pm 0.4$. Within the limitations of the available data, there is 
therefore little evidence for spectral variability, although the parameter constraints are
not particularly tight for the lower flux data.

\begin{figure}
\begin{center}
\hspace*{-0.2cm}
\rotatebox{0}{
{\includegraphics[width=235pt]{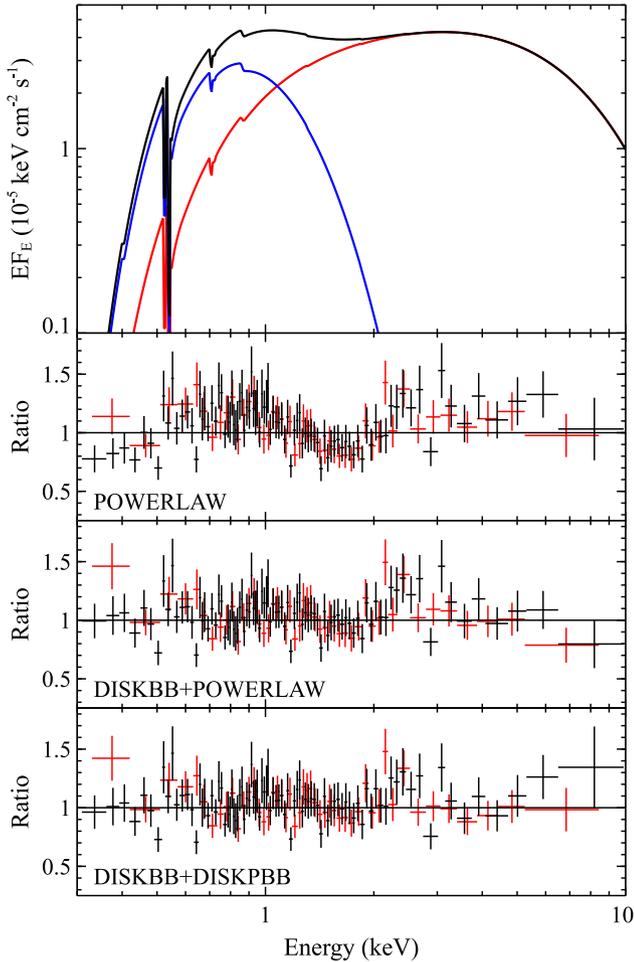}}
}
\end{center}
\vspace*{-0.2cm}
\caption{
\textit{Top panel:} The relative contributions of the best-fit \textsc{diskbb+diskpbb}
model to the high-flux \xmm\ data for NGC\,3044 ULX1. The total model is shown in
black, the \textsc{diskbb} component in blue and the \textsc{diskpbb} component
in red, respectively. \textit{Lower panels:} The data/model ratio for a simple absorbed
powerlaw continuum model, a \textsc{diskbb}+\textsc{powerlaw} continuum and the
\textsc{diskbb+diskpbb} models, respectively. For the ratio panels, the colours have
the same meanings as in Figure \ref{fig_ngc3044_spec}, and the data have again
been rebinned for visual purposes.}
\label{fig_ngc3044_fits}
\end{figure}

Although the absorbed powerlaw model captures the overall shape of the spectrum in
the 0.3-10.0\,keV band fairly well, the high-flux data have sufficient S/N that systematic
residuals to this simple model can be seen (see Figure \ref{fig_ngc3044_fits}), implying
that a more complex continuum model is required. Indeed, the quality of fit provided by
the absorbed powerlaw model for the high flux data is \chisq\ = 353 for 285 degrees of
freedom (DoF), which is not formally an acceptable fit. This residual structure is fairly
typical for extreme ULXs when fit with a single powerlaw model (\eg\
\citealt{Stobbart06, Gladstone09, Mukherjee15}), and indicates the need for distinct
continuum components above and below $\sim$1--2\,keV. We therefore fit the
high-flux data with a few more  complex models often used to describe ULX spectra.
First, we fit a model consisting of a lower energy accretion disc component, and a
higher energy powerlaw continuum. We use the \diskbb\ model (\citealt{diskbb}) for
the former, which implicitly assumes a thin disc profile (\citealt{Shakura73}), such that
the model broadly represents the classic disc--corona geometry seen in sub-Eddington
X-ray binaries. This provides a significant improvement to the simpler powerlaw fit, with 
\chisq/DoF = 284/283. The best-fit parameters are given in Table
\ref{tab_ngc3044_param}.

\begin{table*}
  \caption{Key parameters obtained for the various continuum model fits to the high-flux data available for NGC\,3044 ULX1}
\begin{center}
\begin{tabular}{c c c c c c}
\hline
\hline
\\[-0.2cm]
Model & Parameter & & Continuum Model \\
\\[-0.35cm]
Component & & \textsc{powerlaw} & \diskbb+\textsc{powerlaw} & \diskbb+\diskpbb\ \\
\\[-0.25cm]
\hline
\hline
\\[-0.1cm]
\tbabs\ & $N_{\rm{H}}$ [$10^{21}~\rm{cm}^{-2}$] & $2.1 \pm 0.2$ & $3.5^{+0.7}_{-0.6}$ & $3.0 \pm 0.6$ \\
\\[-0.15cm]
\diskbb\ & $T_{\rm{in}}$ [keV] & -- & $0.16 \pm 0.02$ & $0.19 \pm 0.03$ \\
\\[-0.25cm]
& Norm & -- & $37^{+80}_{-25}$ & $13^{+26}_{-7}$ \\
\\[-0.15cm]
\textsc{powerlaw} & $\Gamma$ & $2.41 \pm 0.07$ & $2.2 \pm 0.1$ & -- \\
\\[-0.25cm]
& Norm [$10^{-5}$] & $6.1 \pm 0.3$ & $5.0^{+0.6}_{-0.5}$ & -- \\
\\[-0.15cm]
\diskpbb\ & $T_{\rm{in}}$ [keV] & -- & -- & $1.7^{+0.7}_{-0.4}$ \\
\\[-0.25cm]
& $p$ & -- & -- & $0.56^{+0.15}_{-0.05}$ \\
\\[-0.25cm]
& Norm [$10^{-4}$] & -- & -- & $3.6^{+5.7}_{-0.5}$ \\
\\[-0.2cm]
\hline
\\[-0.2cm]
$\chi^{2}$/DoF & & 353/285 & 284/283 & 273/282 \\
\\[-0.25cm]
\hline
\hline
\\[-0.15cm]
\end{tabular}
\label{tab_ngc3044_param}
\end{center}
\end{table*}

There is still a mild hint of curvature in the spectrum at higher energies though ($E >
2$\,keV; Figure \ref{fig_ngc3044_fits}). This is seen in the majority of high S/N ULX
spectra, initially implied by \xmm\ (\eg\ \citealt{Gladstone09, Walton4517}) and then
unambiguously confirmed by the higher energy coverage provided by \nustar\ (\eg\
\citealt{Bachetti13, Walton14hoIX, Rana15}). These broadband observations find that
ULX spectra are primarily described by two thermal components below 10 keV. We
therefore also fit a second model that is often considered for ULXs, combining two
accretion disc components.\footnote{The best \nustar\ data available for ULXs
shows that a third continuum component is typically required above $\sim$10\,keV
to model the broadband spectra (\eg\ \citealt{Walton18ulxBB}). However, without
coverage of these energies we cannot say anything about the presence of this
component here, and so restrict ourselves to a simpler two-component model for
the data below 10\,keV.} Specifically, for the higher energy emission we replace the
powerlaw component with a \diskpbb\ model (\citealt{diskpbb}). This allows for the
radial temperature index ($p$) to be an additional free parameter, such that the
model can mimic a thick, advection-dominated super-Eddington accretion disc
(which would be expected to have $p < 0.75$ instead of the $p = 0.75$ appropriate
for thin accretion discs; \citealt{Abram88}). The \diskbb+\diskpbb\ does provide a
moderate additional improvement in fit over the \diskbb\ and powerlaw combination,
with \chisq/DoF = 273/282 (\ie an improvement of $\Delta\chi^{2} = 11$ for one extra
free parameter). As our best-fit model, we show the relative contributions of the
different model components in Figure \ref{fig_ngc3044_fits}, and the parameter
constraints are also given in Table \ref{tab_ngc3044_param}.

\begin{figure}
\begin{center}
\hspace*{-0.2cm}
\rotatebox{0}{
{\includegraphics[width=235pt]{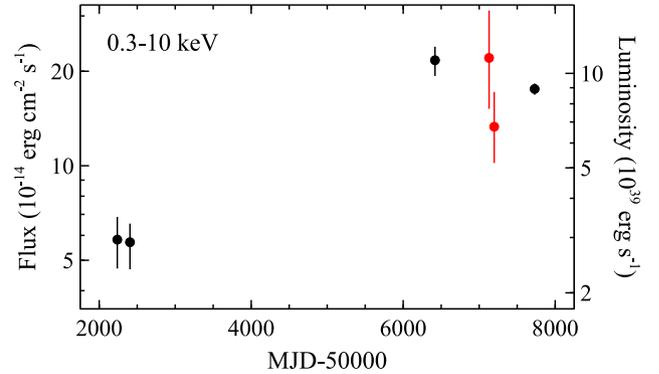}}
}
\end{center}
\vspace*{-0.3cm}
\caption{
Longtime lightcurve for NGC\,3044 ULX1 based on the available X-ray data. \xmm\ 
data are shown in black, and \swift\ in red. Note that the sets of \swift\ observations
taken in April and in June have been combined together here (see Section
\ref{sec_spec}).}
\label{fig_ngc3044_lc}
\end{figure}

The best-fit spectral form for NGC\,3044 ULX1 differs somewhat from that used to
compute the fluxes in 4XMM-DR10, so we also re-calculate the observed 0.3--10\,keV
fluxes for the individual \xmm\ observations using the spectral models for the high-
and low-flux states discussed above. To further examine the long-term behaviour of
the source we also consider the \swift\ data at this stage. These observations can
themselves be split into two main groups, taken in April and June 2015. We process
the combined data from these two sets of observations, compute the average count
rates for each of the two groups using the same 20$''$ regions as for the \xmm\ data
(correcting appropriately for the point spread function), and convert these to fluxes
using the spectral shape implied by the simple powerlaw fits to the \xmm\ data. The
long term lightcurve combining the \xmm\ and \swift\ data is shown in Figure
\ref{fig_ngc3044_lc}. The coverage is admittedly sparse, but the \swift\ fluxes are
consistent with the more recent \xmm\ measurements, and so it appears as though
the higher flux state persisted throughout 2013--2016. We also still find the peak
luminosity of the source to be $L_{\rm{X,peak}} \sim 10^{40}$\,\ergps, confirming the
extreme luminosity implied by our analysis of 4XMM-DR10.

\subsection{Timing Analysis}

The longest of the available \xmm\ observations of NGC\,3044 ULX1, OBSID
0782650101, returns a total of $\sim$5000 net source counts with the \epicpn\
detector, roughly comparable to the quality of data used to detect X-ray pulsations
for some of the known ULX pulsars (\eg\ \citealt{Israel17, Rodriguez20}). We
therefore also perform a search for pulsations on this dataset. We focus on the data
from the \epicpn\ detector here as this has both the highest count rate and the best
timing resolution of the EPIC detectors (73.4\,ms in the full-frame mode used for
this observation). 

For this analysis we use the pulsar timing tools included in the \hendrics\ package
(\citealt{bachettiHENDRICSHighENergy2018}). Since the pulse period can evolve
rapidly in ULX pulsars, either due to the secular spin-up driven by the extreme
accretion (\citealt{Fuerst16p13, Israel17, Carpano18, Vasilopoulos18}) or orbital
motion of the neutron star (\citealt{Bachetti14nat, Israel17, Fuerst18, Fuerst21}), we
perform an `accelerated' pulsation search, which considers both the frequency of
the pulsations ($f$) and its first derivative ($\dot{f}$) when searching for any
signals. Specifically, we use the \textsc{HENzsearch} script, which performs the
$Z^2_n$ search originally outlined in \cite{Buccheri83}, and allow for $n = 2$
harmonics in our search (\ie we use the $Z^2_2$ statistic). Based on the properties
of the known ULX pulsars, we focus on the frequency range of 0.01--6.75\,Hz, and
the ``fast'' search option utilized allows for $\dot{f}$ values in the range
$\pm$10$^{-8}$\,\hzps. Unfortunately we did not find any promising pulsation
candidates in this observation.

In the absence of a robust pulsation detection, we estimate an upper limit on the
pulsed fraction any undetected signal could have following the method used in
\cite{Walton21}. In short, we use the \textsc{HENzn2vspf} script, which simulates
datasets using the same GTIs and total number of events as the real data, then
uses rejection sampling to modulate the events with an increasingly strong pulsed
signal (assuming a sinusoidal pulse profile, which is appropriate for ULX pulsars),
and finally calculates the $Z^2_2$ for each pulsed fraction to see how strongly
such pulsations would have been detected. We simulate 100 datasets in order to
determine the pulsed fraction at which the $Z^2_2$ statistic reaches $\sim$40.
This threshold roughly corresponds to a 3$\sigma$ detection, and thus indicates
the equivalent upper limit on the pulsed fraction that could still be present in the
real data. We find an upper limit on the pulsed fraction of $\sim$22\% when
considering the full \xmm\ bandpass.

\subsection{The Nature of NGC\,3044 ULX1}

NGC\,3044 ULX1 is a new extreme ULX discovered in our analysis that already has
high S/N data available in the archive. Although it has always been in the ULX regime
whenever observed with our current X-ray facilities (considering \xmm\ and \swift\ in
combination, we have observations from 6 different epochs), sometime between
2002 and 2013 it seemed to jump up by a factor of $\sim$3--4 in luminosity from
$L_{\rm{X}} \sim 3 \times 10^{39}$\,\ergps\ to $L_{\rm{X}} \sim 10^{40}$\,\ergps,
where it seems to have remained since (see Figure \ref{fig_ngc3044_lc}).

The 0.3--10.0\,keV X-ray spectrum observed during this high-flux period is very
similar to other extreme ULXs: the flux below 10\,keV is dominated by two continuum
components that primarily contribute above and below $\sim$1--2\,keV and,
although it is not a strong statistical detection, there is a hint of spectral curvature in
the higher energy component. We note in particular that, although there is no higher
energy coverage available here, the spectrum of NGC\,3044 ULX1 is highly
reminiscent of that seen from Holmberg II X-1 -- another extreme ULX with
$L_{\rm{X,peak}} \sim 10^{40}$\,\ergps\ -- during the broadband observations
performed with \xmm, \suzaku\ and \nustar\ in 2013 (\citealt{Walton15hoII}). As
noted previously, this was part of a series of broadband observations of ULXs (\eg\ 
\citealt{Bachetti13, Walton14hoIX, Rana15, Mukherjee15}) that robustly confirmed
earlier indications from \xmm\ (\eg\ \citealt{Stobbart06, Gladstone09, Walton4517})
that the high-energy spectra of ULXs are distinct from those seen from
sub-Eddington black holes. While the spectra from these systems are typically
dominated by Comptonisation in an optically-thin `corona' above $\sim$2\,keV
(\eg\ \citealt{Haardt91}), the spectra of ULXs instead seem to be dominated by two
thermal components below 10\,keV, before falling away steeply at higher energies.
Indeed, the best-fitting model for the high-flux \xmm\ spectra from NGC\,3044
ULX1 consists of two thermal accretion disc components. 

The distinct broadband spectra of ULXs, along with the detection of X-ray
pulsations (\citealt{Bachetti14nat, Fuerst16p13, Israel17, Israel17p13, Carpano18,
Sathyaprakash19, Rodriguez20}) and extreme outflows (\citealt{Pinto16nat, Pinto17,
Pinto20, Walton16ufo, Kosec18}) from a growing number of these systems have,
together, helped clearly establish that the majority of the ULX population is
dominated by super-Eddington accretors. In this context, the two continuum
components seen in ULXs below 10\,keV likely represent the complex thermal
emisson from a hot, super-Eddington accretion disc (and potentially its associated
outflow; \eg\ \citealt{Poutanen07, Middleton15}). Given its similarity to other, better
studied ULXs that are now well accepted to be super-Eddington accretors,
NGC\,3044 ULX1 is therefore likely another super-Eddington system. As discussed
above, these sources are of particular interest, as they may provide a local
observational window into the conditions required to rapidly grow SMBHs in the
early universe (\eg\ \citealt{Banados18nat}).

We have searched for X-ray pulsations from NGC\,3044 ULX1, which would
unambiguously identify the accretor as another neutron star and confirm its nature
as a highly super-Eddington system. We focused on the 2016 data, which by far
have the best S/N among the available observations, and searched for pulsations
over the 0.01--6.75\,Hz frequency range based on the properties of the known
ULX pulsars, but unfortunately we did not find a robust detection of any such
variations. However, even though the available data has quite high S/N, we can only
place an upper limit of $\sim$20\% on the pulsed fraction of any pulsations present
during this observation. Although pulsations of the strength seen in NGC\,300 ULX1
can therefore be firmly excluded (pulsed fraction of $\sim$60\% below 10\,keV;
\citealt{Carpano18}), other known ULX pulsars exhibit pulsed fractions that are lower
than this limit in the \xmm\ bandpass (\eg\ \citealt{Sathyaprakash19, Rodriguez20}).
Furthermore, even in ULXs that are known to be pulsars the pulsations can be
transient, and are not always observed even when the data should have sufficient
S/N to see them (\eg\ \citealt{Israel17, Bachetti20}). As such, even though we have
not seen any clear evidence for X-ray pulsations from NGC\,3044 ULX1, we cannot
exclude the possibility that this is another neutron star ULX.

Although the comparison with other ULXs is quite compelling, obtaining higher
energy data would be of particular use in order to more robustly confirm NGC\,3044
ULX1 as another super-Eddington accretor. In the known ULX pulsars the pulsed
fraction is seen to increase with energy, perhaps because non-pulsed components
from the accretion flow make a more significant contribution below $\sim$10\,keV
(\eg\ \citealt{Walton18ulxBB}). Higher energy coverage would therefore help to
mitigate against these issues in terms of further pulsation searches, and would also
allow us to extend the continuum spectroscopy above 10\,keV, and further confirm
that the broadband spectrum of NGC\,3044 ULX1 is similar to other ULXs.
Unfortunately, given both the fairly low peak flux from NGC\,3044 ULX1 ($\sim$2
$\times$ 10$^{-13}$\,\ergpcmsqps\ in the 0.3--10\,keV band; see Figure
\ref{fig_ngc3044_lc}) and the fairly close proximity of SN1983E (see Figure
\ref{fig_ngc3044_img}), meaningful observations of NGC\,3044 ULX1 with \nustar\
would likely be very challenging. This may, instead, be a suitable target for a facility
like the \textit{High Energy X-ray Probe} (\textit{HEX-P}; \citealt{HEXP_tmp}), which
would have both superior sensitivity and imaging capabilities to \nustar.

\section{Summary and Outlook}
\label{sec_conc}

We have compiled a new catalogue of ULX candidates, combining the latest data
releases from each of the \xmm, \swift\ and \chandra\ observatories (the 4XMM-DR10,
2SXPS and CSC2 source catalogues, respectively). Our new catalogue contains
\nulx\ sources residing in \nhostgals\ different host galaxies, making it the largest ULX
catalogue compiled to date. Of these, \nulxnew\ sources are catalogued as ULX
candidates for the first time. Our sample also contains \nhlx\ HLX candidates, of which
\nhlxnew\ are new catalogue entries. We have made significant efforts to clean the
catalogue of known non-ULX contaminants (e.g. foreground stars, background AGN,
supernovae), and estimate that the remaining contribution of unknown contaminants
is $\sim$20\%. Our primary motivation here is to unearth new sources of interest for
detailed follow-up studies, and among this new catalogue we have already found one
new extreme ULX candidate with high S/N data in the archive: NGC\,3044 ULX1. This
shows a factor of at least $\sim$4 variability on long timescales, based on the
available \xmm\ and \swift\ data, with a peak luminosity of $L_{\rm{X,peak}} \sim
10^{40}$\,\ergps\ to date. The \xmm\ spectrum of the source while at this peak flux is
reminiscent of other extreme ULXs (and Holmberg II X-1 in particular), and is best-fit
by a model combining two thermal accretion disc components. This likely indicates
that NGC\,3044 ULX1 is another member of the ULX population accreting at
super-Eddington rates.

We anticipate this new catalogue will be a valuable resource for planning further
observational campaigns, both with our current X-ray imaging facilities (\xmm,
\chandra, \swift, \nustar) and with upcoming missions such as \xrism\ and, in particular,
\athena. Our new catalogue should also help to facilitate further studies of ULXs at
longer wavelengths, particularly in the era of the new optical, NIR and radio facilities
due to come online (the thirty-metre class ground-based observatories, \textit{JWST},
the SKA). Such work will be vital for determining the contribution of ULX pulsars to the
broader ULX population, their accretion physics, the prevalence of extreme outflows
among the ULX population and the impact of the winds launched by super-Eddington
accretors, and for the hunt for the first dynamically confirmed black hole ULX. Further
iterations of the \xmm, \swift\ and \chandra\ serendipitous surveys, combined with the
upcoming results from \erosita\ (\citealt{EROSITA}), will also allow us to continue
expanding this ULX sample in the future.

\section*{ACKNOWLEDGEMENTS}

DJW acknowledges support from the Science and Technology Facilities Council (STFC)
via an Ernest Rutherford Fellowship (ST/N004027/1).
TPR also acknowledges support from STFC via consolidated grant ST/000244/1.
SM acknowledges financial support from the Spanish Ministry MCIU under project
RTI2018-096686-B-C21 (MCIU/AEI/FEDER/UE), cofunded by FEDER funds and from
the Agencia Estatal de Investigaci\'{o}n, Unidad de Excelencia Mar\`{i}a de Maeztu, ref.
MDM-2017-0765.
This research has made use of data obtained with \xmm, an ESA science mission with
instruments and contributions directly funded by ESA Member States, as well as
public data from the \swift\ data archive.
This work has also made use of data obtained from the \chandra\ Source Catalog,
provided by the \chandra\ X-ray Center (CXC) as part of the \chandra\ Data Archive,
as well as public data from the Swift data archive. 
This paper made use of the Whole Sky Database (WSDB) created by Sergey Koposov
and maintained at the Institute of Astronomy, Cambridge by Sergey Koposov, Vasily
Belokurov and Wyn Evans with financial support from STFC and the European
Research Council (ERC), as well as the Q3C software (\citealt{Q3C}).
This research has also made use of the
NASA/IPAC Extragalactic Database (NED), which is funded by the National Aeronautics
and Space Administration and operated by the California Institute of Technology, as
well as the SIMBAD database, operated at CDS, Strasbourg, France, and we further
acknowledge usage of the HyperLEDA database.

\section*{Data Availability}

All of the raw data underlying this article are publicly available from ESA's \xmm\ Science
Archive\footnote{https://www.cosmos.esa.int/web/xmm-newton/xsa}, NASA's HEASARC
database\footnote{https://heasarc.gsfc.nasa.gov/} and NASA's \chandra\ Data
Archive \footnote{https://cxc.harvard.edu/cda/}. The primary X-ray catalogues
(4XMM\footnote{http://xmmssc.irap.omp.eu/Catalogue/4XMM-DR10/4XMM\_DR10.html},
2SXPS\footnote{https://www.swift.ac.uk/2SXPS/}
CSC2\footnote{https://cxc.harvard.edu/csc/}),
galaxy catalogues
(HyperLEDA\footnote{http://leda.univ-lyon1.fr/},
CNG\footnote{https://www.sao.ru/lv/lvgdb/} and
Cosmicflows\footnote{http://edd.ifa.hawaii.edu/})
and general catalogues
(NED\footnote{https://ned.ipac.caltech.edu/},
SIMBAD\footnote{http://simbad.u-strasbg.fr/simbad/})
used in this work are also all publicly available via the links provided. The final catalogues
of ULX candidates produced here will also be made publicly available via the VizieR
archive\footnote{https://vizier.u-strasbg.fr/viz-bin/VizieR} after the publication of this
work.

\bibliographystyle{/Users/dwalton/papers/mnras}

\bibliography{/Users/dwalton/papers/references}

\label{lastpage}

\end{document}